\documentclass[twocolumn]{aa}
\usepackage{graphicx}
\usepackage{graphicx,xcolor}
\usepackage{txfonts}
\usepackage{multirow}
\usepackage{lscape}
\usepackage{longtable}
\usepackage{comment}
\usepackage{placeins}
\usepackage[breaklinks=true]{hyperref}
\usepackage{ulem}
\usepackage{orcidlink}

    \hypersetup{
        colorlinks=true,
        linkcolor=blue,
        filecolor=magenta,      
        urlcolor=blue,
        citecolor=blue
    }
\usepackage{tabularx}
\usepackage{natbib}
\usepackage{soul}

\newcommand{\aminvsg}{$a_{\mathrm{min,\,a-C}}$}
\newcommand{\abvsg}{M$_{\mathrm{a-C}}$/M$_{\mathrm{H}}$}
\newcommand{\lpdr}{$l_{\rm{PDR}}$}

\newcommand{\mum}{$\mu$m}
\newcommand{\cm}{a-C}

\newcommand{\CM}{a-C:H/a-C}
\newcommand{\py}{a-Sil/a-C}

\newcommand{\nmax}{$n_{0}$}

\newcommand{\jwst}{\textit{JWST}}
\newcommand{\jwstmiri}{\textit{JWST}/MIRI}
\newcommand{\jwstnircam}{\textit{JWST}/NIRCam}

\newcommand{\chiTOT}{$\chi_{\mathrm{tot}}^{2}$}

\newcommand{\HII}{\textrm{H~{\textsc{ii}}}}

\usepackage{color}

\usepackage{xcolor}

\definecolor{cbpurple}{rgb}{0.47, 0.37, 0.94}

\begin{document} 
\makeatother

%   \title{JWST observations of the Horsehead photon-dominated region}
    \title{JWST observations of photodissociation regions}

   \subtitle{III. Dust modelling at the illuminated edge of the Horsehead PDR}

\author{
M. Elyajouri\inst{1,2}%\orcidlink{0000-0002-6086-2337}} 
\and
A. Abergel\inst{1} \and
N. Ysard\inst{3,1} \and
E. Habart\inst{1} \and
T. Schirmer\inst{4} \and
A. Jones\inst{1} \and
M. Juvela\inst{5} \and
B. Tabone\inst{1} \and
L. Verstraete\inst{1} \and
K. Misselt\inst{6} \and
K.\,D. Gordon\inst{2} \and
A. Noriega\textendash Crespo\inst{2} \and
P. Guillard\inst{7} \and
A.\,N. Witt\inst{8} \and
M. Baes\inst{9} \and
P. Bouchet\inst{10} \and
B.\,R. Brandl\inst{11} \and
O. Kannavou\inst{1} \and
P. Dell'ova\inst{1} \and
P. Klassen\inst{12} \and
B. Trahin\inst{2} \and
D. Van De Putte\inst{13}
}

\institute{
Institut d'Astrophysique Spatiale (IAS), Université Paris\textendash Saclay, CNRS, Orsay, France \and
Space Telescope Science Institute (STScI), 3700 San Martin Drive, Baltimore, MD 21218, USA \and
Institut de Recherche en Astrophysique et Planétologie (IRAP), Toulouse, France \and
Department of Space, Earth and Environment, Chalmers University of Technology, Onsala Space Observatory, Sweden \and
Department of Physics, University of Helsinki, Finland \and
Steward Observatory, University of Arizona, Tucson, AZ 85721-0065, USA \and
Sorbonne Université, CNRS, Institut d'Astrophysique de Paris (IAP), 98\,bis bd Arago, 75014 Paris, France \and
Ritter Astrophysical Research Center, University of Toledo, Toledo, OH 43606, USA \and
Sterrenkundig Observatorium, Universiteit Gent, Gent, Belgium \and
Université Paris\textendash Saclay, Université Paris Cité, CEA, CNRS, AIM, 91191 Gif\textendash sur\textendash Yvette, France \and
Leiden Observatory, Leiden University, P.O. Box 9513, 2300 RA Leiden, The Netherlands \and
UK Astronomy Technology Centre, Royal Observatory Edinburgh, Blackford Hill, Edinburgh EH9 3HJ, UK \and
Department of Physics \& Astronomy, The University of Western Ontario, London ON N6A 3K7, Canada
}

\date{October 28, 2025}
 
  \abstract
   {The interpretation of infrared measurements of photon-dominated regions (PDRs) relies on understanding the properties of dust. Additionally, the dependence of dust properties on the environment provides key insights into dust composition, evolution, and formation/destruction processes. This work is conducted as part of the Physics and Chemistry of PDR Fronts program dedicated to the study of dust and gas in PDRs with the James Webb Space Telescope (JWST).}
  % aims heading 
   {A significant component of interstellar dust consists of carbonaceous nano-grains   which often dominate the mid-infrared output of PDRs. In this paper, we study the evolution of the nano-grains across the illuminated edge of the Horsehead Nebula and especially their abundance and size properties.}
  % methods heading 
   {We use NIRCam (3.0, 3.35 and 4.3\,$\mu$m) and MIRI (5.6, 7.7, 10.0, 11.3, 12.8, 15.0, 18.0, 21.0 and 25.5\,$\mu$m) photometric bands, along with NIRSpec and MRS spectroscopic observations to map the illuminated edge of the Horsehead.  We model dust emission, including the aromatic and aliphatic infrared (IR) bands, using the THEMIS interstellar dust model together with the 3D radiative transfer code SOC, in order to fit the photometric bands.
   }
  % results heading 
   {A detailed modeling of high angular resolution JWST data ($\sim$ 6 times higher than that of former observations) allows us to obtain quantitative constraints on the size distribution of nano-grains. In addition, original constraints on the optical properties of these nano-grains are derived from the JWST NIRSpec spectroscopic data. We find that the diffuse interstellar medium (DISM) dust cannot account for the observed data, and it is necessary to use evolved grains. A sharp increase in density is observed at the illuminated edge, consistent with recent ALMA observations which reveal a very sharp transition between molecular and ionized gas. Although the PDR length along the line of sight (\lpdr) could not be directly determined from this study, we estimate an upper limit of $\sim$ 0.015 pc based on geometric considerations and low extinction measured in the infrared. This constraint implies a lower limit on the abundance of small grains (\abvsg > 0.003), showing that small grains are not depleted at the external edge of the Horsehead Nebula, unlike in other PDRs like the Orion Bar. 
}
  % conclusions heading 
   {Our findings indicate a high-density environment and a less steep size distribution for nano-grains at the illuminated edge, in contrast with the diffuse ISM. This implies that nano-grain destruction mechanisms, such as UV-induced destruction, might be less efficient in the Horsehead's moderate-UV field than in PDRs with more intense radiation, like the Orion Bar. These results support a model where nano-grain population recovery, potentially through grain reformation due to fragmentation of larger grains, is slower in moderate-UV environments, leading to a unique dust size distribution at the edge of the Horsehead Nebula.}

   \keywords{infrared: ISM -- dust -- extinction / photon-dominated region (PDR) -- ISM: individual objects: Horsehead -- radiative transfer
               }

   \maketitle
\section{Introduction}
Photon-dominated regions (PDRs) are interfaces between HII regions and molecular clouds, irradiated by nearby energetic stars \citep{hollenbach1997a, hollenbach1999,wolfire2022}. These regions, with their varying physical conditions, are ideal for studying dust evolution. Mid-IR spectra of PDRs, analyzed using ISO and Spitzer data, show significant variations due to small grains overlaying the hot dust continuum \citep[e.g.][]{peeters_rich_2002, Peeters04, Rapacioli2006, abergel_isocam_2002, berne2007}. Herschel far-IR data have enabled the study of large grain emission in thermal equilibrium and probed the densest parts of PDRs.
The combination of SPIRE, PACS, and Spitzer maps has provided full emission spectra (3–500\,\mum) of all dust particles that have been compared to radiative transfer models. In NGC 7023, \cite{abergel2010} illustrated the dramatic influence of radiative transfer on spatial structures observed at long wavelengths.
\cite{Arab2012} show that radiative transfer with diffuse-ISM dust alone cannot reproduce the observations of the Orion Bar PDR. 

Using recent James Webb Space Telescope (JWST) observations as part of an Early Release Science program \citep[PDRs4All,][]{Berne2022,Habart2023jwst,peeters2024}, the most detailed dust evolution modeling of the edge of a molecular cloud to date is that of the Orion Bar PDR \citep{Elyajouri2024}. The authors were able, for the first time, to spatially resolve the steep variation of dust emission at the illuminated edge of the Orion Bar PDR. By considering THEMIS model \citep[see][]{Jones_2013,jones2014,jones2017,kohler15,Ysard2015} with carbonaceous nano-grains and submicronic coated silicate grains, they derive unprecedented constraints on the properties of dust across the Orion Bar. Moreover, they find that the nano-grains must be strongly depleted with an abundance (relative to the gas) 15 times less than in the diffuse ISM. The NIRSpec and MRS spectroscopic observations reveal variations in the hydrogenation of the carbon nano-grains. The lowest hydrogenation levels are found in the vicinity of the illuminating stars suggesting photo-processing while more hydrogenated nano-grains are found in the cold and dense molecular region, potentially indicative of larger grains.
However, the Orion Bar is illuminated by an extremely high-UV radiation field \citep{Marconi1998}, with $G_0 \sim 10^4$, whereas $G_0 = 1.7$ represents the average radiation field in the local interstellar medium \citep{Draine1978}. Such intense radiation is not typical of most UV-illuminated molecular gas in the Milky Way and other galaxies, which usually experience lower to moderate radiation fields. Therefore, studying dust evolution in PDRs with a moderate radiation field is crucial for a broader understanding of the physical structures of PDRs and the evolution of the physico-chemical characteristics of the gas and dust with the local conditions (density and illumination).  

The Horsehead Nebula serves as an archetype of moderately excited PDRs. 
Due to its proximity \citep[$\sim 400$~pc,][]{anthony-twarog_h-beta_1982, 2018A&A...616A...1G}, and favored geometry (nearly edge-on), the PDR located at the edge of the Horsehead nebula is an excellent template for low-UV illuminated PDRs  \citep[with $G_0$ $\sim$ 100;][]{abergel_isocam_2003}. 
The Horsehead has been extensively studied, leading to
a large dataset on dust observations \citep{abergel_isocam_2003,Teyssier2004,compiegne_aromatic_2007,pety_are_2005,compiegne2008,Arab2012} 
, gas observations \citep[e.g.][]{habart_2005,goicoechea2006,Gerin2009,Guzman2011,Pety2012,Ohashi2013,LeGal2017} and laboratory experiments on thermal processed and UV-irradiated dust grains analogues  \citep[see][]{smith1984,Zubko2004,Alata2014,Alata2015,Duley2015}. Recent high-angular resolution ($\sim$0.5$\arcsec$) observations of the millimetric molecular emission obtained with ALMA have allowed to reveal a sharp transition (< 650 AU) between the molecular and ionized gas at the edge of the PDR \citep{hernandez-vera2023}.

Observations with Spitzer have refined the observational
picture of the variations of the nano-grains emission among PDRs but did not provide a clear picture of the origin of the variations. \cite{compiegne_aromatic_2007} proposed a scenario in which PAHs survive in \HII~regions and \cite{compiegne2008} interpreted that mid-infrared spectral variations cannot be explained solely by radiative transfer effects and are therefore a consequence of dust evolution across the Horsehead. 

\cite{schirmer2020} fit Spitzer and Herschel observations of the Horsehead PDR with the THEMIS dust evolution model. Despite the limited angular resolution, a radiative transfer model including the evolution of dust grains (size, abundance, properties) allowed them to highlight a strong depletion of nano-grains in the irradiated outer part (i.e.
$\leq$ 0.05 pc or 25$\arcsec$ from the edge) of this PDR and a size distribution shifted towards larger particles compared to the diffuse ISM. 
However, due to the limited angular resolution of Spitzer, no spatial variation of this depletion could be detected along the illuminated edge. Additionally, since attenuation effects were neglected at the time, no definitive conclusions could be drawn. In the denser part, evolved grains (aggregates, with/without ice mantles) were needed to reproduce the observations. Extending to several PDRs, \citet{schirmer2022} report broad PDR-to-PDR variations in dust properties and a common tendency toward nano-grain depletion at UV-exposed interfaces, and propose fragmentation of large grains in radiative pressure–driven collisions as a viable scenario.

\begin{figure*}[tbh]
\includegraphics[width=\textwidth]{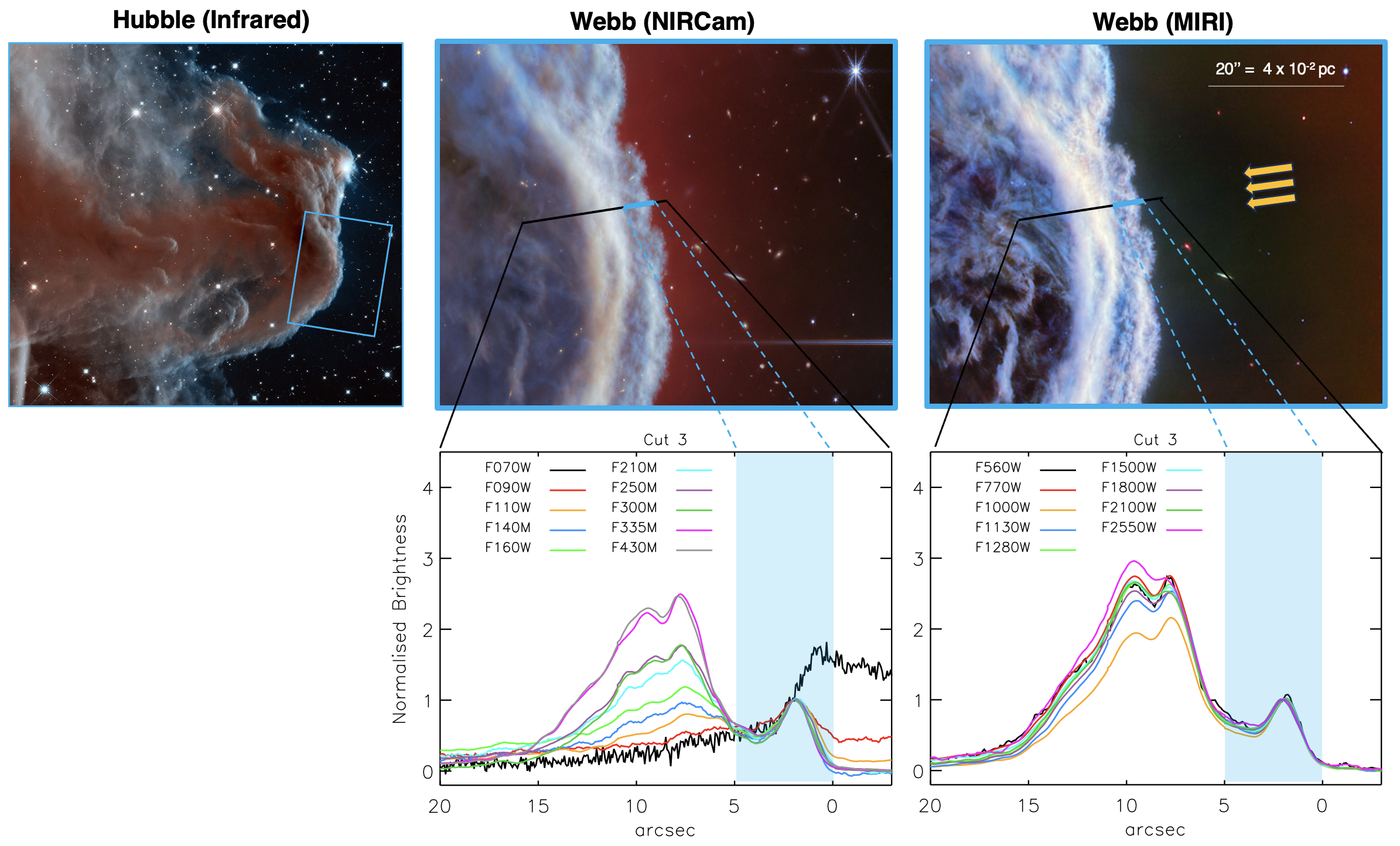}
\caption{\textbf{Top: (Left)} Hubble’s view of the Horsehead Nebula at near-infrared wavelengths of 1.1 \,$\mu$m (blue/cyan) and 1.6 \,$\mu$m (red/orange); NASA, ESA, and the Hubble Heritage Team (STScI/AURA).
\textbf{(Middle)} a zoom-in image of part of the Horsehead nebula as seen by the NIRCam instrument. This image is composed of light at wavelengths of 1.4 and 2.5 \,$\mu$m (blue), 3.0 and 3.23 \,$\mu$m (cyan), 3.35 \,$\mu$m (green), 4.3 \,$\mu$m(yellow), and 4.7 and 4.05 \,$\mu$m (red). \textbf{(Right)} The same zoom-in region visualised by the JWST instrument MIRI. In this image, blue represents light at wavelengths of 5.6, 7.7, and 10 \,$\mu$m ; green is 11, 12, and 15 \,$\mu$m; and red is 18, 21, and 25 \,$\mu$m.  The NIRCam and MIRI images are from the recent ESA/Webb release, NASA, CSA, K. Misselt (University of Arizona) and A. Abergel (IAS/University Paris-Saclay, CNRS).
The blue solid lines
correspond to the cut used in our study.
\textbf{Bottom:} Profiles of the relative brightness (normalized at a distance of 2\arcsec from the edge) for NIRCam and MIRI filters, adapted from \cite{Abergel2024}.} 
\label{fig:HH_maps}
\end{figure*}

Recent JWST observations have now provided unprecedented detail of the spatial variations at the edge of the Horsehead Nebula. As part of the Guaranteed Time Observations (GTO) program (PID 01192), \citet{Abergel2024} used NIRCam and MIRI with 23 filters to explore the global structure of the PDR, resolving its spatial complexity from 0.7 to 25 \mum\, with an angular resolution of 0.1–1$\arcsec$ (Fig.\,\ref{fig:HH_maps}). Their study provided the first quantitative analysis of dust attenuation, suggesting that attenuation could be the dominant process behind the observed large-scale color variations. They found that dust attenuation is non-negligible across most lines of sight through the inner regions of the Horsehead, yet a small, localized region at the very edge (0 to 5$\arcsec$) appears to be unaffected by attenuation. 

Motivated by the high angular resolution of the JWST and the need to address the previously overlooked attenuation effects, this study focuses specifically on the clean, non-attenuated external filament. We aim to constrain the properties of nano-grains at the illuminated edge of the Horsehead PDR. Using spectral and photometric data from the JWST, we seek to understand the mechanisms driving nano-grain formation and destruction in this region. This approach will enable us to constrain the optical properties of nano-grains, building on similar work by \citet{Elyajouri2024} in the Orion Bar.

This paper is structured as follows: 
Sect.\,\ref{sec:obs} presents the observations of the Horsehead. In Sect.\,\ref{sec:rt}, we describe the radiative transfer modeling of the dust emission, whereas in Sect.\,\ref{sec:methodo_free}
we derive constraints on their properties and the physical conditions of the PDR. In
Sect.\,\ref{main_results}, we describe the PDR models that best reproduce the observations. We discuss the evidence of a steep density gradient
and dust evolution in the Horsehead in Sect.\,\ref{sec:discussion}, and we summarize the results and conclusions in Sect.\,\ref{sec:conclusion}.

\section{Data\label{sec:obs}}

The observations are part of the JWST Guaranteed Time Observing (GTO) "Physics and Chemistry of PDR Fronts" program 1192.

\subsection{Imaging}
We use observations in 12 photometric bands (3.0, 3.35 and 4.3~\mum) for \jwstnircam~and (5.6, 7.7, 10.0, 11.3, 12.8, 15.0, 18.0, 21.0 and 25.5~\mum) for \jwstmiri~(see Fig.~\ref{fig:HH_maps}). The detailed processing of these \jwst~maps is described in \citet{Abergel2024}. Calibration uncertainties are estimated at 4\% for \jwstnircam~bands \citep{Abergel2024} and 3\% for \jwstmiri~filters \citep{Gordon2025}.
We study the observed emission profiles through a perpendicular cut (labeled “2” in \citealt{schirmer_2020} and “3” in \citealt{Abergel2024}). This cut corresponds to the location of the IFU observations (see Sect.\ref{data:spectra}).

\subsection{Spectra\label{data:spectra}}
We also utilize the spectral extractions from \citet{Misselt2025}; the full details of
the observations, data reduction, region definition and spectral extraction for the NIRSpec IFU and MIRI/MRS data are given therein. In the present work, we focus on
the DF1 region spectrum which probes the externally illuminated edge of the PDR. This region provides the most stringent constraints on the aromatic and aliphatic C–H features and spatially coincides with the external filament located 0–5\arcsec\ from the illuminated edge (see Fig.~\ref{fig:HH_maps}), which is unaffected by dust attenuation along the line of sight as presented in \citet{Abergel2024}.

\subsection{Bands and line contamination}
Using NIRSpec and MRS spectra, \citet{Misselt2025} quantified contamination by H$_2$ lines for each JWST band. 
Although most of the flux in the photometric bands arises from dust emission, several filters are significantly contaminated by H$_2$ emission lines (as seen in the NIRSpec spectrum in Fig.~\ref{fig:bandgap}). 
For DF1, the contamination reaches about 50\% in F300M, 20\% in F430M and F1000W, and 10\% in F335M.  At shorter wavelengths (1–2.5~\mum), the broad NIRCam filters are dominated by gas emission lines. We therefore exclude these bands from our analysis. 

We correct all photometric bands for H$_2$ contamination using the line-contribution values in \citet[][Table~3]{Misselt2025}. After this correction, the 3.35~\mum~filter primarily captures continuum emission and the aromatic C-H stretch band at 3.3~\mum, along with the aliphatic band at 3.4~\mum. The 7.7, 11.3 and 12.8~\mum\, filters trace the broad emission bands commonly referred to as aromatic infrared bands (AIBs), which can serve as a proxy for the ionization state of Polycyclic Aromatic Hydrocarbons  \citep[PAHs; e.g.][]{Joblin96_ngc1333, Galliano:08}. Meanwhile, the filters at 5.6, 10.0, 15.0, 18.0, 21.0, and 25.5~\mum\, are dominated by continuum emission, analogous to the corresponding bands in WISE, Spitzer, and IRAS.

\section{Radiative Transfer modelling\label{sec:rt}}

Photon-Dominated Regions (PDRs) are dense environments exposed to strong radiation fields, which leads to differences in the optical properties and composition of dust compared to diffuse regions. Our objective is to identify, characterize, and quantify these variations in dust properties. However, radiative transfer effects play a critical role in shaping the observed dust emission, making it necessary to use detailed modeling to accurately recover the grain properties in regions like the Horsehead PDR.

We adopted the methodology detailed in \cite{Elyajouri2024}, combining a radiative transfer method with the THEMIS dust model \cite{Jones_2013}. The radiative transfer calculations were performed using the SOC code \citep{juvela2019}, which requires input on the radiation field, density profile, geometry, and dust model within the PDR. 

\subsection{PDR structure and density profile\label{subsec:structure}}
\begin{figure}
\includegraphics[width=\linewidth]{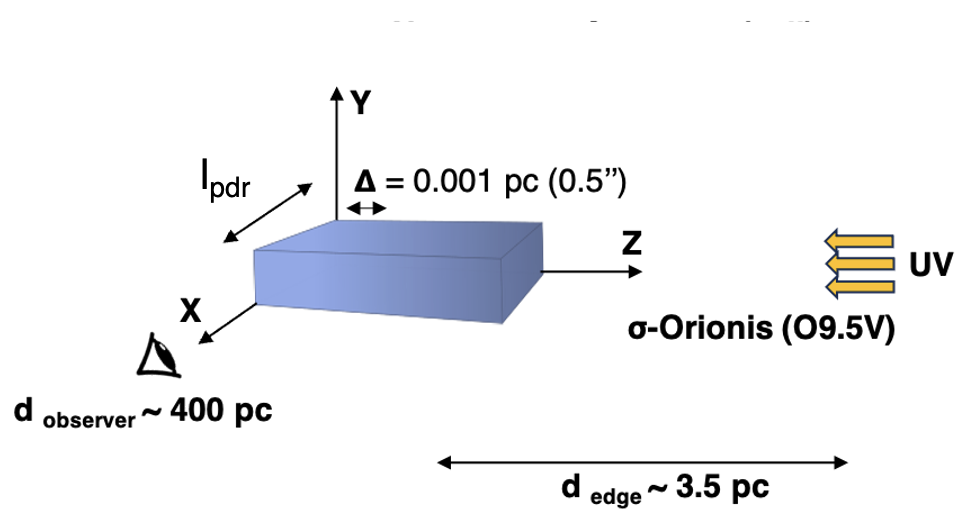}  
    \caption{Schematic illustration of the Horsehead PDR illuminated by a radiation field (from the right). $l_\text{PDR}$ parameter is the length of the PDR along the line of sight. Refer to Table \ref{soc_setup} for the corresponding values.} \label{fig:structure}
\end{figure}

The Horsehead nebula is illuminated by the O9.5V binary system $\sigma$-Orionis \citep{warren_photometric_1977}, which has an effective temperature of $T_{\mathrm{eff}}\sim$~34600~K \citep{schaerer_combined_1997}. This star system is located approximately $d_{\mathrm{edge}}\sim$~3.5~pc from the edge of the PDR.  
In this model, the PDR is represented by a plane-parallel rectilinear slab that is externally illuminated by a single blackbody source. This idealized geometry assumes an edge-on PDR where photons enter perpendicularly to the PDR edge, with no inclination effects, at a projected distance of \(d_{\text{edge}} = 3.5\) pc. The radiation field parameters (the radius of the
exciting star, its temperature, and its distance from the cloud),
are derived from stellar parameters of the O9.5V binary system $\sigma$-Orionis \citep{warren_photometric_1977} and fixed to give an illumination equal to 100 $\times$ $G_0$ in Habing units. The input parameters of our model  %values for the cloud and point source 
are given in Table~\ref{soc_setup} and the geometry illustrated in Fig.~\ref{fig:structure}.

\begin{table}
\caption{SOC radiative transfer inputs.}
\label{soc_setup}
\begin{tabular}{lc}
\hline\hline
Name & Values \\
\hline
\multicolumn{2}{c}{Cloud Geometry} \\
\hline
model size & 0.05 $\times$ 0.02 $\times$ 0.01 pc \\
 & 25\arcsec $\times$ 10 \arcsec$\times$ 5 \arcsec \\
\hline
\multicolumn{2}{c}{Star} \\
\hline
location (x,y,z) & (0 pc, 0 pc, 3.5 pc) \\
temperature & 34\,600 K \\
$G_{0}$ & 100 \\
\hline
\multicolumn{2}{c}{Observer} \\
\hline
Distance & 400 pc \\
%\lpdr & 0.05 pc  (25 \arcsec)\\
\hline
\end{tabular}
\end{table}   

\begin{figure}
\centering    
\includegraphics[width=0.8\linewidth]{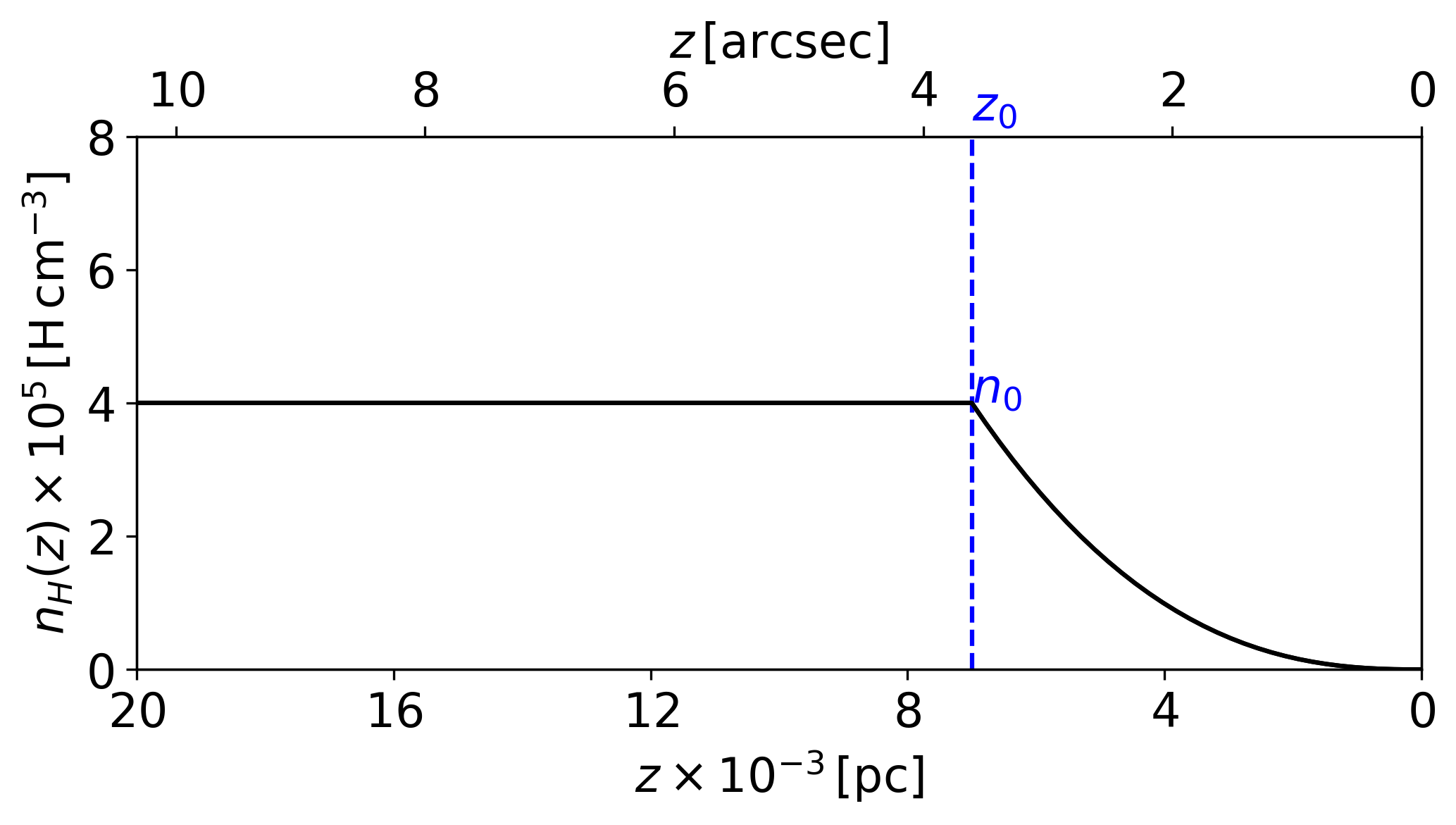}  
\caption{Top: Assumed density profile across the illuminated edge of the Horsehead PDR (see Sect.~\ref{subsec:structure}).
} \label{fig:density}
\end{figure}

The density profile is set by the previous gas and dust models of PDR edges \citep{habart05,Elyajouri2024} along the z-direction:

\begin{equation}
    \label{eq:density_profile}
    n_{\text{H}}(z)=\left\{ 
\begin{array}{l l}
  n_{0} \left(\frac{z}{z_{0}}\right)^{\gamma}  & \quad \text{if $z<z_{0}$}\\
  n_{0} & \quad \text{if $z>z_{0}$ ,}\\ \end{array} \right.
\end{equation}
where $z$ is the distance from the edge of the PDR, the power-law index $\gamma$ governs the steepness of the density front and is equal to 2.5. The maximum value of the density $n_{0}$ is reached at the depth $z_{0}$ and is assumed to remain constant for $z>z_{0}$. The density profile is illustrated in  Fig.~\ref{fig:density}.  

In the following, we define the length of the PDR along the line of sight as \lpdr, it is related to the column density $N_{\rm H, tot}(z)$
by $N_{\rm H, tot}(z)$ = n$_{H}$(z) $\times$ \(l_{PDR}\), where n$_{H}$ is the density along the line of sight. 
The parameters  \lpdr, \nmax, and \(z_{0}\) are treated as free parameters.

\subsection{Dust model\label{subsec:themis}}
The dust grains have the properties from the original version of THEMIS \citep[The Heterogeneous dust Evolution Model for Interstellar Solids,][and references therein]{jones2017}. THEMIS is based on observational constraints and laboratory measurements on interstellar dust analogs that are amorphous hydrocarbons, a-C(:H) \citep[e.g.,][]{jones2012a,jones2012b,jones2012c} and amorphous silicates, a-Sil. The THEMIS model for the diffuse ISM (DISM) is therefore composed of three dust populations that are built upon these two materials:
\begin{itemize}
    \item \textit{Carbon nano-grains (a-C):}
    this population follows a power-law size distribution with an exponential cut-off. 
    These nano-grains are small, partially hydrogenated, (sub-)nanometric a-C carbon grains (0.4 $\leq$ a $\leq$ 20 nm, hereafter, a-C nano-grains). Similar to polycyclic aromatic hydrocarbons (PAHs), they contain intrinsic aromatic domain sub-structures, but also olefinic and aliphatic sub-structures.  
    \item \textit{Large Carbon Core-mantle grains (\CM):} this population follows a log-normal size distribution and consists predominantly of a-C:H/a-C core-mantle grains (99\% of the mass), with a small fraction of pure a-C. %{\bf Ant, delete grains}. 
    We refer to this population as \CM~grains.
    \item \textit{Carbon coated silicate grains (a-Sil/a-C):} This population consists of amorphous silicate grains surrounded by an aromatic carbonaceous a-C mantle following a log-normal size distribution.
\end{itemize}

\begin{figure}[]
	\includegraphics[width=\linewidth]{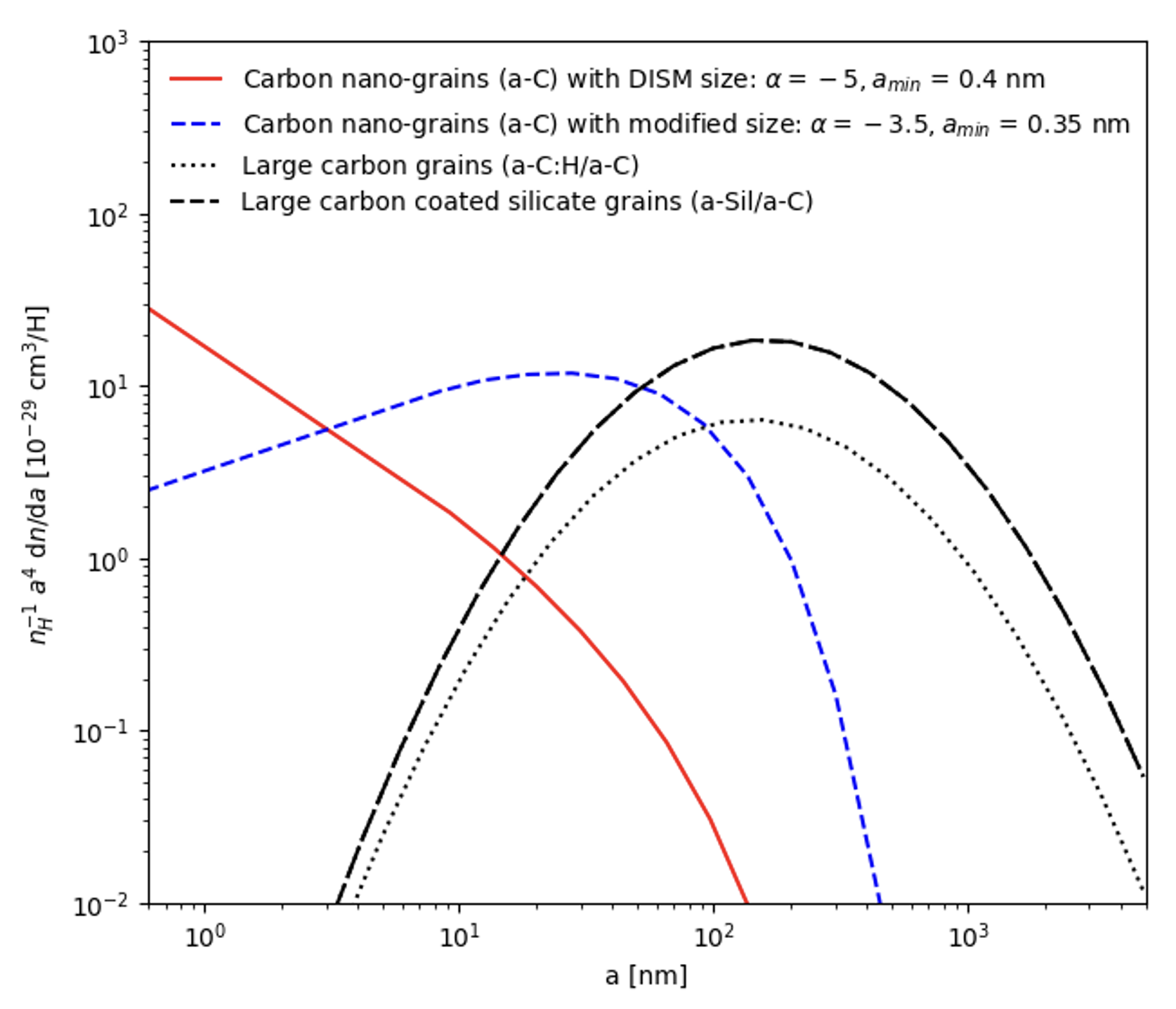}
    \caption{Size distributions of the dust mixtures from THEMIS using the dust emission tool DustEM \citep{compiegne2011}
    (parameters are listed in Table.\,\ref{tab:size_distribution}). The red, black dotted line, and black dashed line 
    correspond to \cm, a-C:H/a-C and a-Sil/a-C respectively. The Blue dashed line corresponds to the distribution of a-C grains for the best-fit parameters at the edge of the Horsehead.
    }
    \label{fig:s_dist}
\end{figure}

The size distributions of these three populations are shown in
Fig.\,\ref{fig:s_dist} (red and black lines) and the associated parameters are detailed in Table.\,\ref{tab:size_distribution}.\\
 
In the near-IR (1 to 5 $\mu m$) and mid-IR (5 to 30 $\mu m$), dust emission comes mainly from the a-C nano-grains. To study the influence of dust evolution on the emission across the Horsehead, we use a-C nano-grains with modified size distributions. Three parameters associated with their size distribution are varied: 1) the abundance, that is, the a-C to gas mass ratio, \abvsg; 2) the minimum size, \aminvsg; and 3) the exponent of the power-law size distribution, $\alpha$.

\begin{table}
\caption{\label{tab:size_distribution} Size distribution parameters
for the dust population (see Sect.\,\ref{subsec:themis}). p-law is a power-law with an exponential tail and log-n is 
a log-normal distribution.}
\centering
\begin{tabular}[h]{lccccccc}
    \hline\hline 
     Name & size & $\alpha$ & $a_{\mathrm{min}}$ & $a_{\mathrm{max}}$ & 
     $a_{\mathrm{c}}$ & $a_{\mathrm{t}}$ & $a_{0}$  \\
     \hline 
     & &  & [nm] & [nm] & [nm] & [nm] & [nm] \\ 
     \hline 
     \cm & p-law & 5 & 0.4 & 4900 & 10 & 50 & - \\
     \CM & log-n & - & 0.5 & 4900 & - & - & 7 \\
     \py & log-n & - & 1 & 4900 & - & - & 8 \\
     \hline
     \hline 
\end{tabular}
\end{table}
%revision 2
We note that we employ the original version of THEMIS rather than the recently updated THEMIS 2.0 \citep{Ysard2024}. This choice is motivated by two considerations: First, the near- and mid-infrared observations analyzed in this work are dominated by emission from carbonaceous nano-grains, whose optical properties remain unchanged between THEMIS versions. The updates in THEMIS 2.0 primarily concern silicate optical constants, which may affect the fit in the 9-10~$\mu$m region (as discussed in Sect.~\ref{sub:nirspec_soc}) but do not significantly impact our constraints on the carbonaceous nano-grain population that dominates the observed spectral range. Second, adopting the same dust model used in recent comparable studies of the Horsehead and Orion Bar PDRs \citep{schirmer2020,schirmer2022, Elyajouri2024} ensures that the variations we report reflect genuine physical differences between these environments rather than artifacts of model version changes. Future work will incorporate THEMIS 2.0 to evaluate potential improvements in the silicate emission region.

To compare our dust emission models with JWST observations, we applied post-processing to the model outputs. This process includes convolving the models with the point spread functions (PSFs) of NIRCam and MIRI, obtained from WebbPSF \citep{Perrin20214}, and integrating them over the relevant photometric bands using the photon-to-electron conversion efficiency.

\section{ Methodology}\label{sec:methodo_free}

As demonstrated in \cite{Elyajouri2024}, combining both spectroscopic and photometric approaches yields a more complete understanding of dust behavior in PDRs than relying on either method alone. By constraining both the large spatial-scale imaging data and the spectral data limited to a narrow cut, which overlaps with a small part of the imaging data, we can better understand the dust properties at various spatial scales and validate our model with the limited spectral data.

To begin with, we compare the DustEM spectral energy distributions (SEDs), calculated under optically thin conditions without radiative transfer, with NIRSpec data. This allows us to determine the hydrogenation state of the nano-grains using the 3.3-to-3.4~$\mu$m band ratio. We then compare the photometric data obtained from NIRCam and MIRI, with model-generated profiles using THEMIS and SOC.  Subsequently, using the parameters derived from these photometric comparisons, we extract the SOC-computed spectrum at peak emission and validate it against the NIRSpec spectrum.
This approach allows us to thoroughly investigate how variations in parameters such as dust grain size, composition, radiation field, and density affect the emission profiles. A grid-based analysis is used to systematically explore a range of free parameters.

The free parameters in our model\footnote{The model parameters are only free within the sense that a given set of coupled parameters (e.g. minimum size and band gap) determine the observable dust properties over a wide wavelength range (e.g. NIR to MIR or MIR to millimeter). One cannot therefore adopt arbitrary parameter values.} incorporate elements associated with both the density profile and nano-grain properties, as well as the structure of the PDR: 

\begin{enumerate}
    \item \emph{Position ($z_0$) and density ($n_0$)} are the density profile parameters already defined in (Eq.\ref{eq:density_profile}). They influence the spatial emission profiles (steepness and width) for all of the NIRCam and MIRI filter fitting.
    \item \emph{Band gap energy ($E_\text{g}$) - hydrogenation state}: is completely determined by the electronic structure of a-C(:H) nano-grains and is primarily and strongly constrained by the spectral data. This parameter particularly affects the 3.3-to-3.4~$\mu$m band ratio and as such is the initial focus of our grid exploration using NIRSpec spectra.

    \item \emph{Abundance ($M_\text{a-C}$/$M_\text{H}$)}: The a-C dust mass to gas ratio, represents the abundance of the small a-C nano-grains ($\sim$ sub 20 nm radius) relative to the gas.
    \item \emph{Minimum size ($a_\text{min,a-C}$)}: the smallest a-C nano-grain size in the distribution. This parameter impacts both the spectral shape and emission continuum slope.
    \item \emph{Slope ($\alpha$)}: of the a-C power-law size distribution. Like $a_\text{min,a-C}$, this plays a significant role in determining the spectral shape and emission continuum slope.  
    \item \emph{Length of the PDR ($l_\text{PDR}$)} along the line of sight. We assume that the \lpdr~does not affect the shape of the dust spectrum and is considered as a multiplying factor on the dust spectrum (assuming $l_\text{PDR}$ $\leq$ 0.05 pc). The intensity increases linearly with $ l_\text{PDR}$ without altering the shape of the dust spectrum \citep[e.g.][]{schirmer2020}.
\end{enumerate}

In our approach, we constrain each set of parameters sequentially using specific observational tracers. This methodology is designed to be time-efficient and reduce the number of free parameters at each step. The order chosen to constrain the free parameters is critical, as it ensures physical consistency between spectroscopy and imaging data. Our procedure reflects extensive testing with various parameter sequences, and the final parameters are obtained from joint optimization over all observables, informed by the methodology developed in previous PDR studies \citep{schirmer2020,schirmer2022, Elyajouri2024}.

\section{Main results\label{main_results}}
The sequential approach proceeds as follows: (1) We first use NIRSpec/MRS spectroscopy to fix the band gap energy $E_g$ from the 3.4/3.3~$\mu$m band ratio (Sect.~\ref{sec:eg}). This spectroscopic constraint is crucial because it also narrows the viable ranges of $\alpha$ and $a_{\rm min}$---parameter combinations that might fit the photometric bands but contradict the spectral continuum shape and slope are immediately rejected. (2) We then constrain the density profile parameters ($n_0$, $z_0$) using the Full Width at Half Maximum (FWHM) of the emission profiles (Sect.~\ref{sec:density}). As demonstrated in Fig.~\ref{fig:fwhm_vs_dust}, the FWHM is primarily sensitive to the density parameters and remains largely unaffected by dust properties ($a_{\rm min}$, $\alpha$, $M_{\rm a-C}/M_{\rm H}$). This allows us to determine $n_0$ and $z_0$ independently of the grain size distribution parameters. (3) Finally, with $E_g$ constrained by spectroscopy and ($n_0$, $z_0$) by the FWHM analysis, we use radiative transfer modeling to constrain the remaining dust parameters ($M_{\rm a-C}/M_{\rm H}$, $a_{\rm min}$, $\alpha$) through grid exploration (Sect.~\ref{constrain_prop}).

The choice of this specific order is essential. Exploring the ($\alpha$, $a_{\rm min}$) parameter space from imaging data alone could yield solutions that appear acceptable in $\chi^2$ space but are inconsistent with the spectroscopic constraints once those are applied. Starting with spectroscopy eliminates such non-physical minima and significantly reduces the viable parameter space.
\subsection{First step: constraining the hydrogenation state \(E_g\) \label{sec:eg}}
We start by constraining the hydrogenation state, $E_\text{g}$, using as a tracer the 3.3/3.4 \mum~ratio derived from the NIRSpec spectrum and following the same approach as in \cite{Elyajouri2024}. Changes in $E_\text{g}$ are most prominently reflected in this ratio, where a decrease (increase) in the band gap corresponds to the removal (addition) of hydrogen atoms from the grain structure, leading to a reduction (enhancement) in the 3.4 µm aliphatic band \citep[see also][]{jones2012c,Jones_2013}.

\begin{figure}
    \centering
    \includegraphics[width=0.5\textwidth]{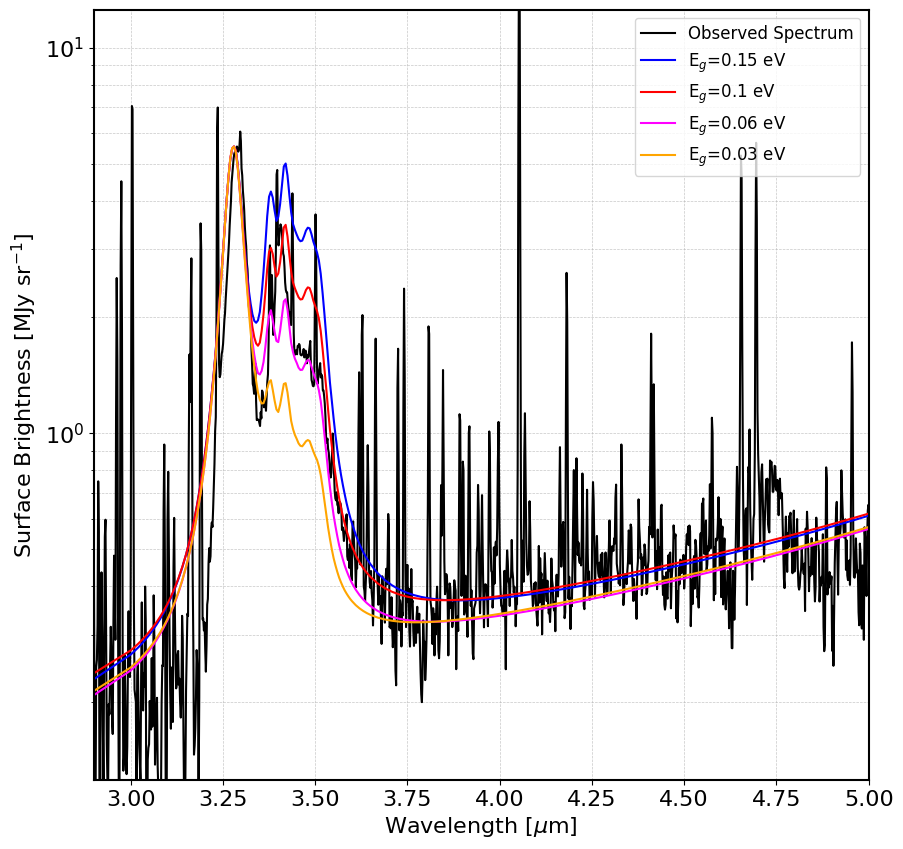}
    \caption{
    Comparison of observed surface brightness and DustEM models normalized to the 3.3~$\mu$m observed band (without radiative transfer) in the illuminated edge of the Horsehead PDR. The black data points represent observations from NIRSpec, while the coloured lines show model predictions with varying band gap energies. The dust parameters used are those of the DISM model, \aminvsg = 0.4 nm, $\alpha$ = -5.}\label{fig:bandgap}
\end{figure}

Fig.~\ref{fig:bandgap} shows the comparison between the modeled SED ($E_\text{g}$ varies from 0.03 to 0.15~eV) and the NIRSpec spectrum corresponding to the illuminated edge of the Horsehead. All SEDs correspond to a minimum grain size of 0.4~nm and $\alpha$ = -5 (Diffuse ISM values). Although the continuum fits relatively well, we note variations of the ratio between the 3.3~$\mu$m aromatic band and the 3.4~$\mu$m aliphatic band. For E$_{g}$ = 0.15 eV, the 3.4~$\mu$m band is significantly overestimated (factor of $\sim$2), while for lower values of the band gap (\(E_\text{g}\) = 0.03 and 0.06 eV), the 3.4~$\mu$m band is significantly underestimated (factor of $\sim$2). 
Regarding the DISM THEMIS model (a 0.1~eV band gap), the fit is reasonably\footnote{Here, we have chosen to adjust only the main bands at 3.3 and 3.4~$\mu$m and the continuum, and to ignore the sub-band around 3.5~$\mu$m to avoid over-interpreting the data. In light of recent spectroscopic observations, it appears that the model, adjusted in 2012 based on observations of the Galactic Center, needs to be updated (band exact positions, widths and strengths)}. good for the continuum and the bands. Therefore, we can determine the best band gap as 0.1 eV.

\subsection{Second step: constraining \(n_0\) and \(z_0\)\label{sec:density}}

Next, we constrain the parameters of the density profile \(n_0\) and \(z_0\) using the Full Width at Half Maximum (FWHM) of the emission profiles as a tracer. This approach is justified because the FWHM is primarily influenced by the parameters \(n_0\) and \(z_0\), while it remains largely unaffected by dust properties. As demonstrated in Fig.~\ref{fig:fwhm_vs_dust}, an increase in \(n_0\) or a decrease in \(z_0\) results in a narrower FWHM (see Fig.~\ref{fig:fwhm_vs_dust}, lower panels). In contrast, dust properties such as \abvsg, \aminvsg, and \(\alpha\) have minimal to no effect on the FWHM (top panels of Fig.~\ref{fig:fwhm_vs_dust} ).

\begin{figure*}
\centering     
\includegraphics[width=0.35\linewidth]{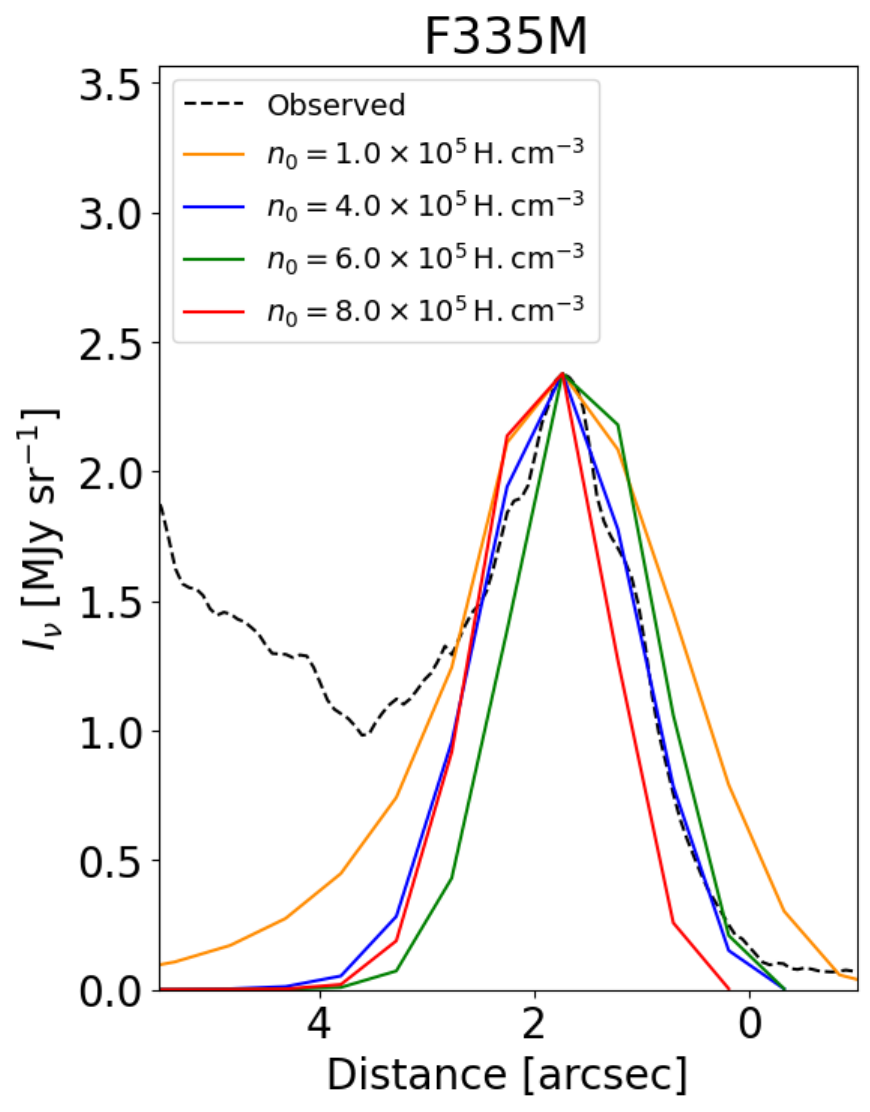}  
\includegraphics[width=0.35\linewidth]{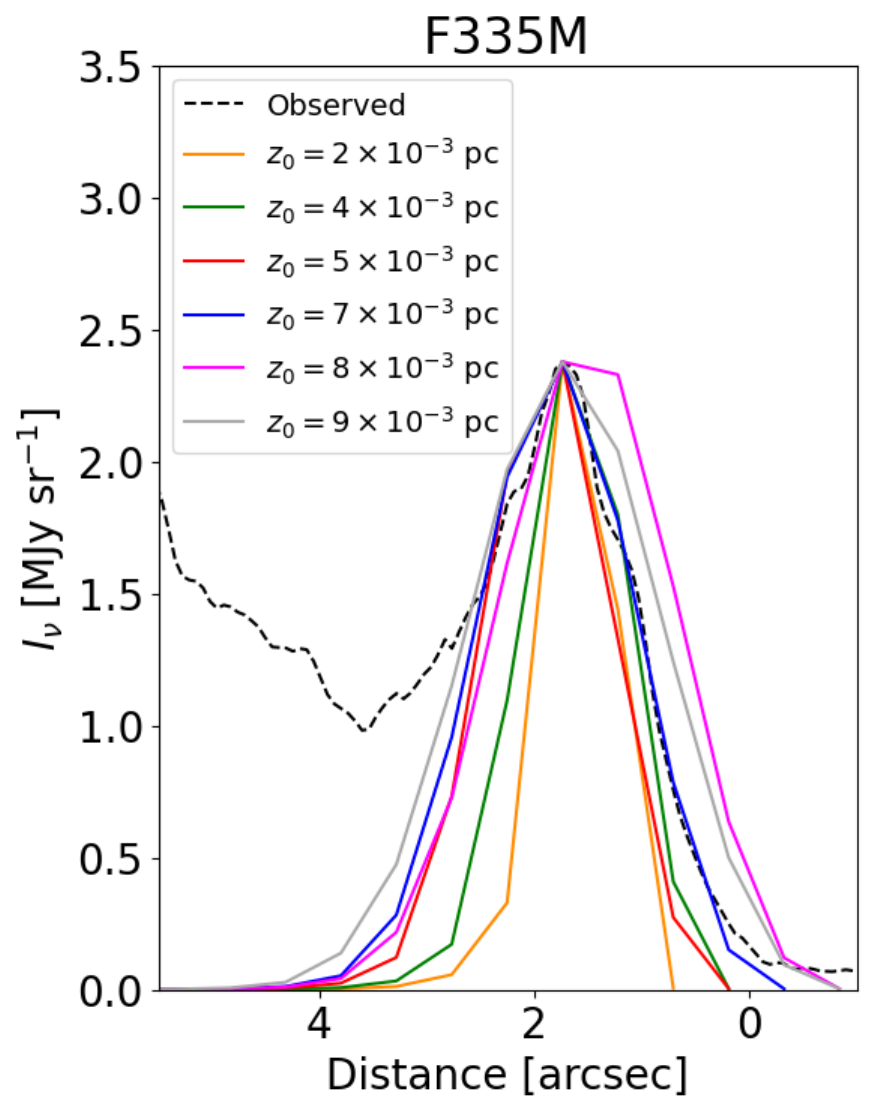}  
\caption{Observed brightness profile (black) in the 3.35 \mum~photometric band, compared with modeled profiles (colored) for varying density parameters \(n_0\) (from \(10^{5}\) to 8 \(\times 10^{5}\) H.cm\(^{-3}\), left) with $z_{0}$ fixed to $7 \times 10^{-3}$ pc and \(z_0\) (from 2 \(\times 10^{-3}\) to 9 \(\times 10^{-3}\) pc, right) with $n_{0}$ fixed to $4 \times 10^{5}$ H cm$^{-3}$). The modelling was performed using the following dust properties: \abvsg \( = 2.3 \times 10^{-3}\), \aminvsg\, = 0.53 nm, and \(\alpha = -4\). The incident radiation field, illuminating the Horsehead nebula from the right, corresponds to a blackbody at 34\,600 K with \(G_0 = 100\). All emission profiles have been convolved with the PSFs and normalized to the observed filament.} \label{fig:variation_density}
\end{figure*}
Figure~\ref{fig:variation_density} presents a comparison between the observed brightness profiles and models with varying density parameters. Since the effect is consistent across all bands, we illustrate this only for the 3.35~\mum~photometric band. In the left panel of Fig.~\ref{fig:variation_density}, we observe that as \( n_{0} \) decreases, the FWHM of the profile broadens. The model with \( n_{0} = 4 \times 10^5 \, \mathrm{H\,cm^{-3}} \) (blue profile) provides the closest match to the observed profile. In the second panel, we vary \( z_{0} \) and find that \( z_{0} = 7 \times 10^{-3} \, \mathrm{pc} \) (in blue) most accurately reproduces the profile shape, in particular the width and the steepness. Another couples of (\(n_{0}, z_{0}\)) with lower \( z_{0} \)  and higher \( n_{0} \)  could reproduce the width but not the steepness.  

While the density value of \( n_{0} = 4 \times 10^5 \, \mathrm{H\,cm^{-3}} \) may initially appear high for the region analyzed here, it is consistent with ALMA observations presented by \citet{hernandez-vera2023}.

\citet{Abergel2024} compare the ALMA observations with the JWST brightness profiles at distances of 0 to 5\arcsec~from the illuminated edge (see their Fig. 12, low panel).

A notable increase in CO emission as well as faint HCO$^+$ emission is observed from 1 to 4 \arcsec from the edge. %an angular resolution of 0.5\arcsec. 
The HCO$^+$ \textit{J} = 4-3 line emission, which effectively traces dense molecular gas (density in the range $10^4-10^6$cm\,$^{-3}$, see \citealt{2016Natur.537..207G}), shows a marked increase immediately behind the illuminated edge. 

The uncertainties on \(n_0\) and \(z_0\) are estimated by assessing the range of parameter values that still provide a good match to the observed FWHM and profile shape. 
Therefore, the parameters that are compatible with the observations and that describe the density profile across the Horsehead are: 

\begin{equation}     \label{eq:density_parameters}
    n_{0} = (4\, \pm \,  1) \times 10^{5}  \,\mathrm{H\,cm^{-3}} \quad ; \quad 
    z_{0} = (7\, \pm \, 1) \times \, 10^{-3}  \,\mathrm{pc} \quad 
\end{equation}

\subsection{Third step: constraining nano-grain properties and \lpdr\label{constrain_prop}}

At this stage, \abvsg, \aminvsg, $\alpha$, and \lpdr~are the only remaining unconstrained parameters and are thus used to run the grid of models. Unlike \abvsg, \aminvsg, and $\alpha$, which vary directly within the grid, \( l_{\text{PDR}} \) is treated as a scaling factor because changes in \( l_{\text{PDR}} \) primarily impact the overall intensity of the dust spectrum, not its shape (see Sect.~\ref{sec:methodo_free}). 
We therefore fix \( l_{\text{PDR}} \) at 0.05~pc, avoiding the need to rerun the grid for different \( l_{\text{PDR}} \) values. This value is comparable to the width of the observed filaments along the cut (25\arcsec) and provides sufficient resolution along the line of sight in the radiative transfer calculations. 

\subsubsection{Method}
We vary the remaining three parameters in the grid as follows:

\begin{enumerate}
    \item \abvsg~ranges from 0.001 $\times$ $10^{-2}$ to 0.4 $\times$ $10^{-2}$  on a logarithmic grid of 44 points, including the DISM standard value of 0.17$\times10^{-2}$.
    \item \aminvsg~varies from 0.35~nm to 0.95~nm in steps of 0.03~nm, including the DISM standard value of 0.4~nm.
    \item $\alpha$ ranges from -10 to -3 in steps of 0.5, including the DISM value of -5.
\end{enumerate}

The maximum value for \abvsg~is set to be $\sim$ 2.3 times the standard value in the diffuse interstellar medium (DISM: 0.17$\times$ $10^{-2}$), while the minimum value is chosen to account for depletion values reported by \cite{schirmer_2020}. The ranges for \aminvsg~and $\alpha$ are broad enough to encompass all physically plausible values, though extreme values can be excluded based on the NIRSpec spectrum, limiting the valid range to those near the DISM values. 

To compare the observed and modeled dust emission profiles, we minimize the following \chiTOT~distribution:
\begin{equation}
    \chi^{2}_{\mathrm{tot}} = \sum_{i\,\in\,\mathrm{filters}} \left(\frac{I_{\mathrm{obs},i}^\text{max} - I_{\mathrm{mod},i}^\text{max}(l_{\text{PDR}})}{\sigma_{i}} \right)^{2} \;,
\end{equation}
where \( I_{\mathrm{obs},i}^{\,\text{max}} \) and \( I_{\mathrm{mod},i}^\text{max}(l_{\text{PDR}})\) are the \( i \)-th band maximum intensities, respectively obtained from the observed and modeled brightness profiles at position $z$, $I_{\mathrm{obs},\,i}(z)$ and $I_{\mathrm{mod},\,i}(z,l_\text{PDR})$ corrected for H$_2$ line contamination as detailed in Sect.~\ref{sec:obs}. We note here that $I_{\mathrm{mod},\,i}(z,l_\text{PDR})$ is proportional to $l_{\rm PDR}$, the PDR length chosen in the model.\\

\noindent The observed band profile is thus matched for a PDR length $l_\text{PDR}^{\,\text{max}}$ fulfilling the constraint
\begin{equation}
    I_{\mathrm{obs},i}^{\,\text{max}} = I_{\mathrm{mod},i}^{\,\mathrm{max}}(l_\text{PDR}) \times\frac{l_{\text{PDR}}^{\,\text{max}}}{l_{\text{PDR}}}.
\end{equation}
For comparison to observations, the model intensity profile is consequently rescaled as follows 
\begin{equation}
    I_{\mathrm{mod},i}(z) = I_{\mathrm{mod},i}(z,l_\text{PDR}) \times\frac{l_{\text{PDR}}^{\,\text{max}}}{l_{\text{PDR}}}
\end{equation}
where \( l_{\text{PDR}} \) is varied logarithmically from $10^{-4}$ to $5 \times 10^{-2}$~pc on a grid of 1000 points to find $l_{\text{PDR}}^{\,\text{max}}$. For simplicity, we will note below $l_{\text{PDR}}$ as the best-fitting (or minimum $\chi^2_\text{tot}$) value instead of 
$l_{\text{PDR}}^{\,\text{max}}$.\\

\noindent The uncertainty \(\sigma_{i}\) for each observed band is given by:
\begin{equation}
    \sigma_{i} = I_{\mathrm{obs},i}^\text{max} \times \sqrt{\delta_{\mathrm{obs},i}^2 + \delta_{\mathrm{H_2},i}^2} \;,
\end{equation}

\noindent where \(\delta_{\mathrm{obs},i} = 4\%\) is the relative calibration error for each band, and \(\delta_{\mathrm{H_2},i} = 10\%\) represents the uncertainty from H$_2$ contamination. \\

\begin{figure}
    \centering
       \includegraphics[width=\linewidth]{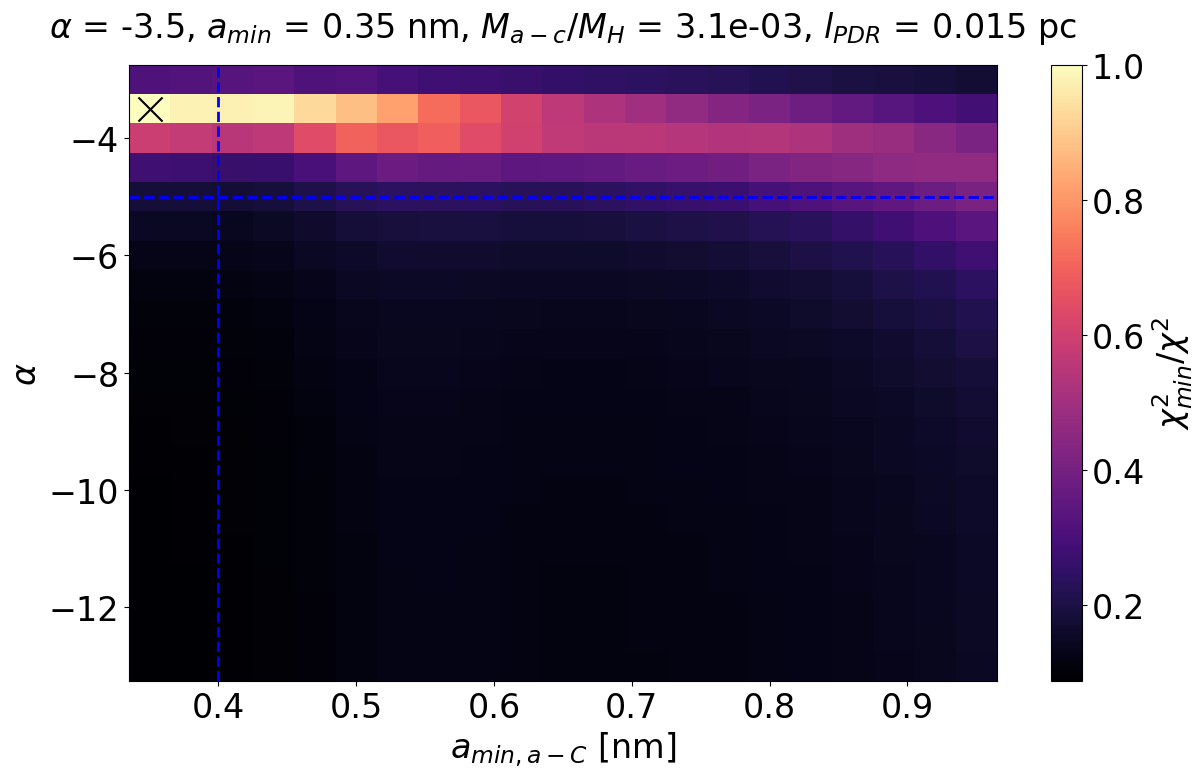}
       \includegraphics[width=\linewidth]{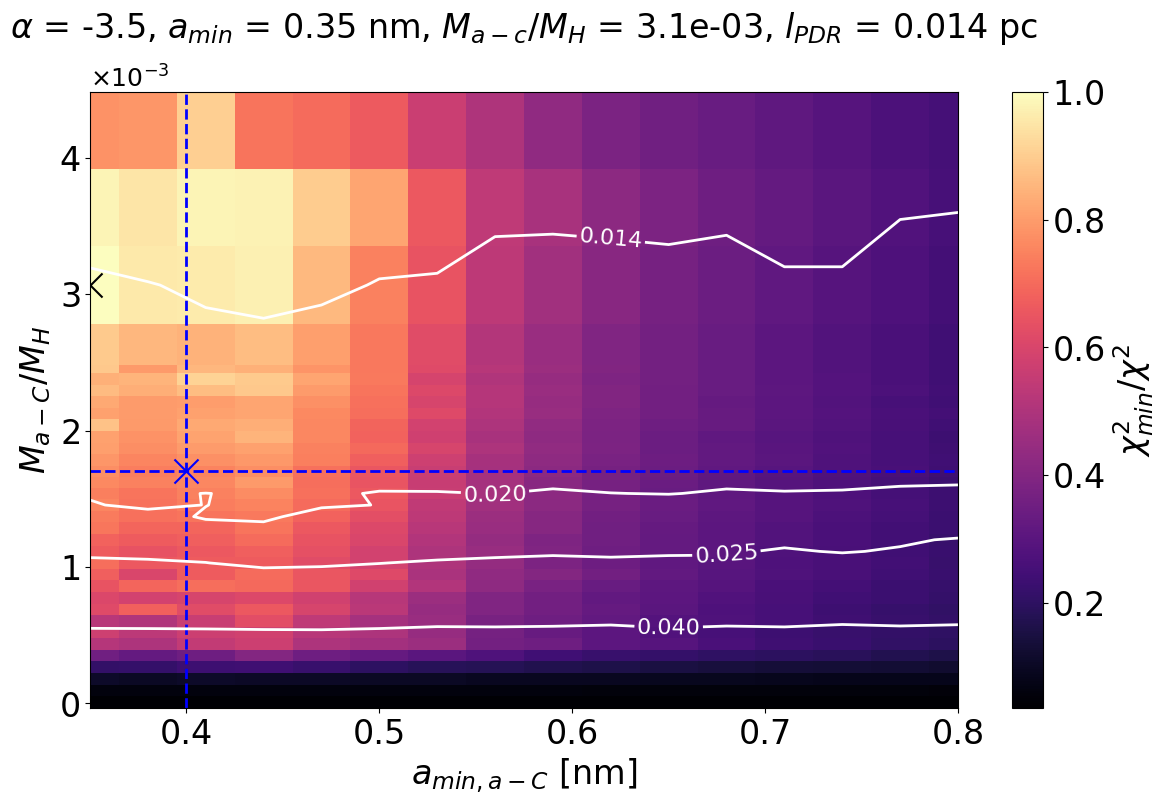}
     \caption{Constraints obtained from $\chi^2_{min}/\chi^2_{tot}$ maps with $\chi^2_{min}$ = min (\chiTOT) and $\chi^2_{min}/\chi^2_{tot}$ = 1 for the best fit. Top: $\chi^2_{min}/\chi^2_{tot}$ in the 2D space ($\alpha$ and \aminvsg). Bottom: $\chi^2_{min}/\chi^2_{tot}$ in the 2D space (\abvsg, \aminvsg).  The white lines indicate the best-fitting \lpdr~for each combination of \abvsg~and \aminvsg. A degeneracy is generally observed between solutions with high \abvsg~and low \lpdr. The cross marker shows the parameters that minimize \chiTOT, representing the best fit. The dashed blue lines indicate the DISM values (\abvsg~= 0.17 $\times 10^{-2}$, \aminvsg~= 0.4 nm, $\alpha$~= -5).
     }\label{fig:chi2}
\end{figure}

\subsubsection{Result}

We present the \(\chi^2\) distribution in two 2D parameter spaces in Fig.~\ref{fig:chi2} with $\chi^2_{min}$ = min (\chiTOT) and $\chi^2_{min}/\chi^2_{tot}$ = 1 for the best fit. The top panel shows the ratio $\chi^2_{min}/\chi^2_{tot}$ as a function of \(\alpha\) and \(a_{\text{min}}\), revealing a minimum around \(\alpha = -3.5\) and \(a_{\text{min}} = 0.35\) nm. 
The color gradient indicates that moving away from these values increases \(\chi_{\text{tot}}^2\), which corresponds to a poorer fit. It appears that the degeneracy in the ($\alpha, a_{\text{min}}$) parameter space is limited. 

The bottom panel of Fig.~\ref{fig:chi2} displays $\chi^2_{min}/\chi^2_{tot}$ in the space of \abvsg~and \(a_{\text{min}}\) for a fixed \(\alpha = -3.5\).  For \(a_{\text{min}} = [0.35-0.45]\)\,nm, \(\chi_{\text{tot}}^2\) reaches a minimum regardless of the \abvsg~value. The clear degeneracy we observe here is between \abvsg~and \(l_{\text{PDR}}\). 
The \lpdr~value associated with the minimum of \chiTOT, decreases with an increase in \abvsg. A decrease in \lpdr~reduces dust emission in the near and mid-IR and this effect is counterbalanced by an increase in the value of \abvsg~associated
with the \chiTOT~minimum value in the 2D-space (\abvsg, \aminvsg).  

\begin{figure}
    \centering
       \includegraphics[width=0.5\textwidth]{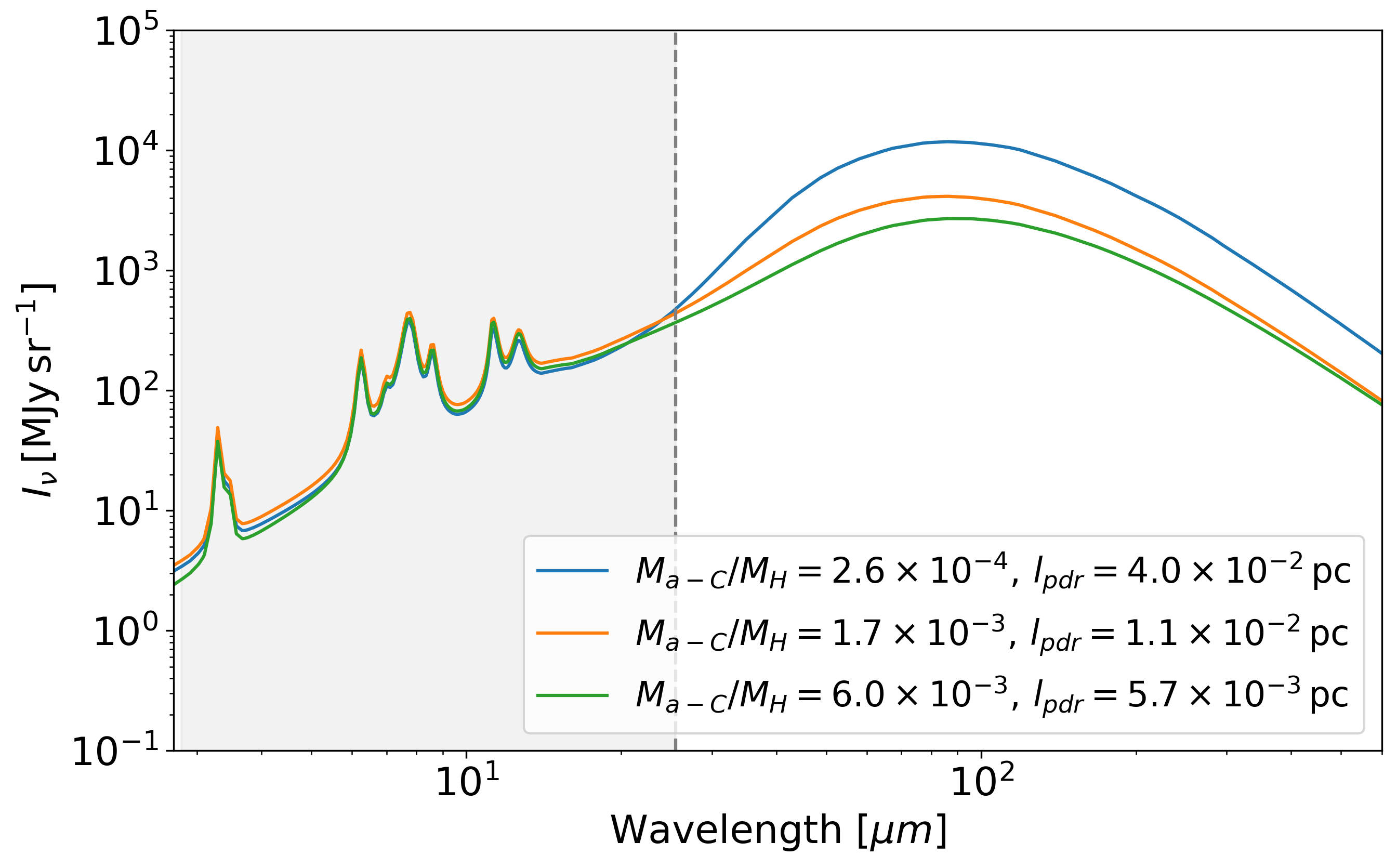}
        \caption{Spectra computed with SOC for different dust abundance values \( M_{\text{a-C}}/M_{\text{H}} \) and Horsehead PDR layer thickness \( l_{\text{PDR}} \), highlighting the degeneracy between dust abundance and \( l_{\text{PDR}} \). The models shown correspond to a strongly depleted abundance (\(2.6\times10^{-4}\), blue), a diffuse ISM-like abundance (\(1.7\times10^{-3}\), green), and a high abundance (\(6.0\times10^{-3}\), orange). An increase in \( M_{\text{a-C}}/M_{\text{H}} \) is counterbalanced by a decrease in \( l_{\text{PDR}} \). The dashed vertical line marks the longest observed wavelength of 25.5\,$\mu$m, which is thus most sensitive to the emission from large grains.  
        }
\label{fig:spec-all}
\end{figure}

The degeneracy between \abvsg~and \(l_{\text{PDR}}\) is expected as we have no observations taken at longer wavelengths and with a comparable angular resolution to independently constrain  \(l_{\text{PDR}}\) using the larger grain emission that is at thermal equilibrium. This can be surprising given the results of \cite{Elyajouri2024}, who showed that large grains produce significant emission at 25~\mum~in the Orion bar. However, the weaker incident radiation field in the Horsehead leads to cooler large grains that emit only beyond the 25~\mum. This is further illustrated in Fig.\,\ref{fig:spec-all}. 

Finally, Even though the value of \lpdr~cannot be determined from 
the data we are using, its upper limit can be independently constrained. First, \lpdr~is likely comparable to the observed width of the filament, so \lpdr~  $\lesssim$ 0.005-0.015 pc ($\lesssim$ 5\arcsec). Second, the low extinction observed in the infrared across this filament imposes that for \( n_{0} = 4 \times 10^{5}\,\mathrm{H.cm^{-3}} \) and $N_{H}/A_{V}$=1.85 $\times 10^{21}$ cm$^{-2}$/mag, \lpdr = \(N_{H}\)/\( n_{0}\) < 0.0075 pc to maintain A$_{V}$ < 5 \cite{Abergel2024}. Consequently, if we have an upper limit on \lpdr~around 0.015 pc, we obtain a lower limit on the abundance of small grains: \abvsg~> 0.003 as shown in Fig.~\ref{fig:chi2}, indicating that the small grains are not depleted in the Horsehead, unlike in the Orion Bar. 
% revision2
Finally, while we cannot rigorously exclude all possible solution degeneracies, as is typical for spectral energy distribution modeling, the combined spectroscopic and photometric approach yields a sharply peaked $\chi^2$ distribution in the ($\alpha$, $a_{\rm min}$) parameter space once $E_g$ is fixed. The residual degeneracy between $M_{\rm a-C}/M_{\rm H}$ and $l_{\rm PDR}$ is further constrained by the independently measured visual extinction and validated through comparison of the radiative transfer spectrum with the combined NIRSpec+MRS observations (Sect.~\ref{sub:nirspec_soc}). Given the current observational constraints and modeling framework, the solution presented represents the only parameter set fully consistent with both the spectroscopic and photometric data. This conclusion will be rigorously reviewed and potentially refined in future studies with extended wavelength coverage.

\begin{table*}[]
    \centering     
     \caption{Overview of the physical parameters defining irradiation ($G_0$, $T_{\mathrm{eff}}$) and density profiles ($n_0$, $z_0$), along with the modeled dust properties (\abvsg, \aminvsg, $\alpha$) for three regions: the edge of the Horsehead PDR (for both this work and in \citealt{schirmer2020}), the Orion Bar \citep[atomic PDR;][]{Elyajouri2024}, and the diffuse interstellar medium \citep[DISM;][]{Jones_2013}. Note that \cite{schirmer_2020} focused on the bright filament, whereas our study probes the illuminated edge of the Horsehead PDR (a spatial scale inaccessible to Spitzer; see Sect.~\ref{discussion:amin})}
    \label{tab:resume2}
    \begin{tabular}{lcccccccc}
        \hline
        \hline
        PDR & $G_0$ & $T_{\mathrm{eff}}$ & \abvsg & \aminvsg & $\alpha$ & ($n_0$ , $z_0$) & \lpdr \\
        & & [\,K\,]  &  & [\,nm\,] & & [10$^{5}$ \,H\,cm$^{-3}$\,, 10$^{-3}$\,pc\,] & [\,pc\,]\\
         \hline
          \multicolumn{8}{l}{Horsehead} \\
          \hline
          This work  & 100 & 35000  & $\geq$ $0.3\times 10^{-2}$ & $0.4\pm 0.05$ & -3.5 & $( 4\, \pm \,  1 ; 7\, \pm \, 1)$ %, $(3 , 8)$, $(5 , 6)$ 
          & $\leq 1.5 \times 10^{-2}$ \\
         Schirmer et al.  & 100 & 35000  & $(0.02\pm 0.01)\times 10^{-2}$ & $0.77\pm 0.03$ & $-6\pm 0.5$ & (2 , 60) & 0.25 \\
\hline
          Orion Bar & 26000 & 38000  & $0.01\times 10^{-2}$ & $\leq 0.5$ &  -5 & (0.9 , $3 $) &  0.12 \\
        \hline
        Diffuse ISM & 1  & - & $0.17\times10^{-2}$ & 0.4 & -5 & - & - \\
        \hline
    \end{tabular}
\end{table*}

\subsection{Comparison with NIRCam and MIRI brightness profiles}
We have seen in the previous section that while there is a degeneracy between the best \abvsg~and \lpdr, we can constrain \aminvsg~and $\alpha$. In fact, \aminvsg~is $\sim 0.35-0.45$ nm, i.e. comparable to the diffuse ISM value (0.4 nm) and $\alpha$ is -3.5, i.e. 1.4 times higher than in the diffuse ISM (-5), as illustrated in Fig.\,\ref{fig:s_dist}. 

Let us now compare the dust modeled and observed emissions across the illuminated edge of the Horsehead for the best-fit parameters: \aminvsg\,= 0.35 nm, $\alpha$ = -3.5, \abvsg~= 3.1 $\times 10^{-3}$ and \lpdr = 0.015 pc. In Fig.\,\ref{fig:best_profiles} (top panel), we see that the width of the synthetic profiles reproduces well the observed profiles, which supports our hypothesis that the profile width is largely independent of dust properties and is instead determined primarily by the density parameters.

The dust emission peaks are generally well reproduced within the error
bars, except at 5.6, 7.7, 10.0 \(\mu\)m and 25 \(\mu\)m. The discrepancies observed at these wavelengths will be discussed in Sect.\,\ref{sub:nirspec_soc}, following the comparison with NIRSpec data.

\subsection{Spectral comparison at the peak position\label{sub:nirspec_soc}}

\begin{figure*}
    \centering
    \includegraphics[width=\linewidth]{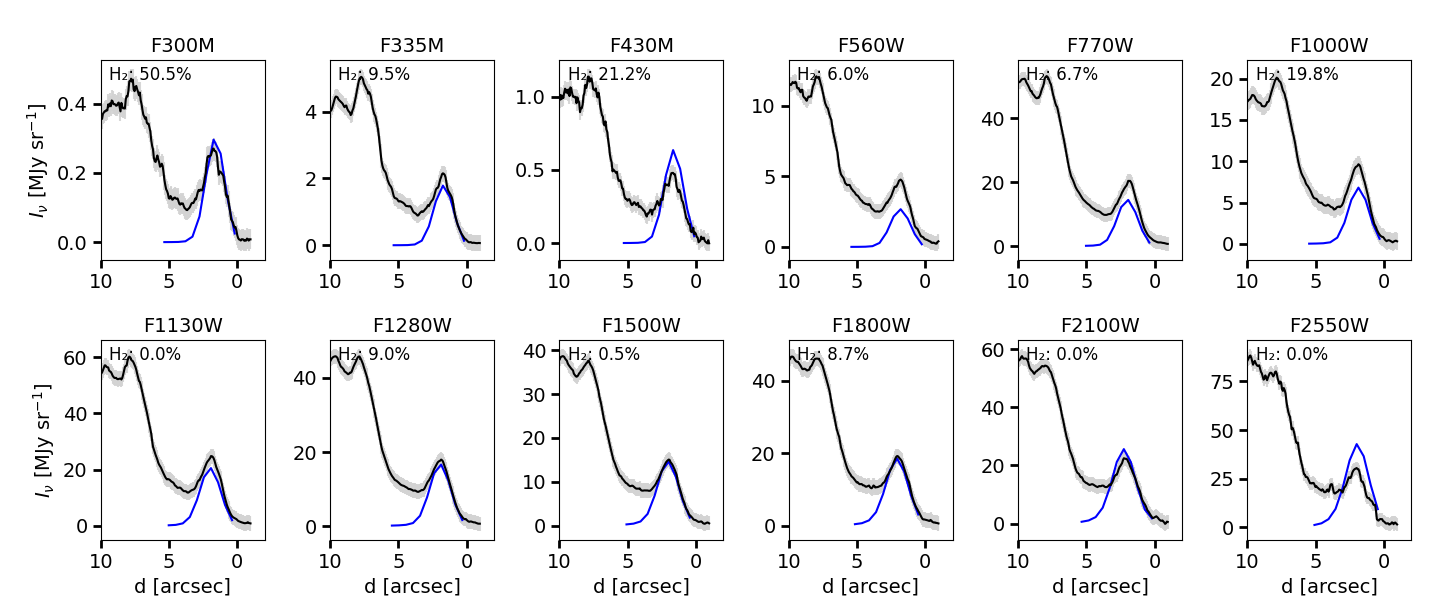}

    \includegraphics[width=\linewidth]{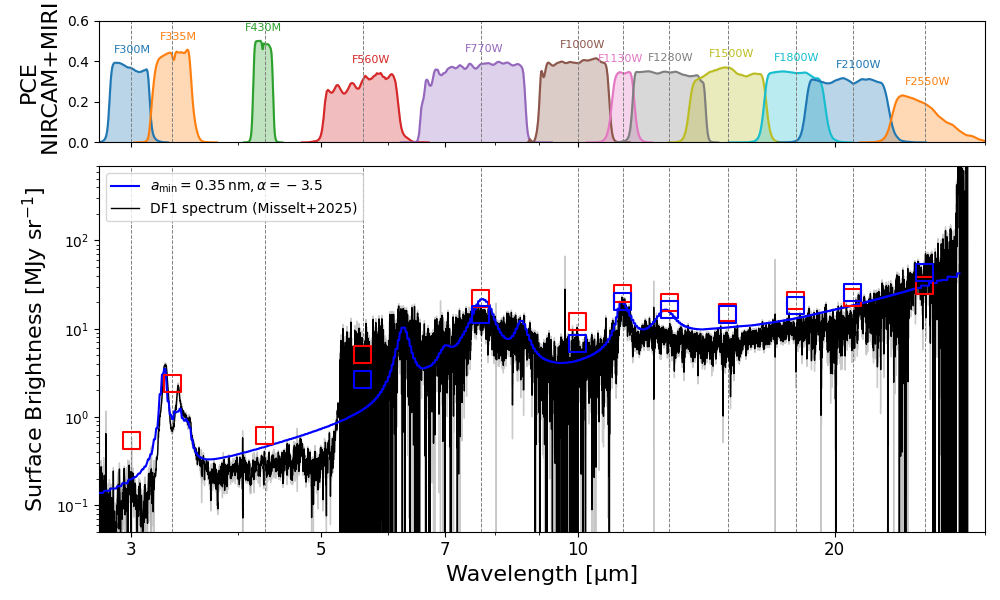} 
    
    \caption{Comparison between JWST photometric and spectroscopic data and the self-coherent models using SOC and THEMIS. 
    Top: Comparison between the observed dust emission (black) and the modeled dust emission (blue) using the best set of dust parameters (from the model that minimized the \chiTOT\, showed in Fig.\,\ref{fig:chi2}) in twelve photometric bands (3.0, 3.3 and 4.8 \mum; NIRCam filters, and 7.7, 10.0, 11.3, 12.8, 15.0, 18.0, 21.0 and 25.5 \mum; MIRI filters). The dust observed emission is shown in black lines. The errors are shown in light gray. The cut considered across the Horsehead is shown in blue in Fig.\,\ref{fig:HH_maps}. Middle: wavelength coverage and photon-to-electron conversion efficiency (PCE) of the NIRCam and MIRI imager filters used in this study. The central wavelengths of the filters are indicated with dashed vertical lines. 
    Bottom: Comparison of the dust emission spectral energy distribution (SED) calculated using our radiative transfer model at the AIB peak emission with the NIRSpec+MRS spectrum for DF1 region of the Horsehead PDR (black). The full extracted spectrum was obtained by subtracting the fitted line contributions from the total flux as presented in \cite{Misselt2025}. The photometric bands maximum values are represented by red squares and the photometric band points from the model are shown in blue squares.}\label{fig:best_profiles}
\end{figure*}

Using the best parameters from the radiative transfer model (\aminvsg\,= 0.35 nm, $\alpha$ = -3.5, \abvsg~= 3.1 $\times 10^{-3}$ and \lpdr = 0.015 pc), we extract the computed spectrum at a distance of 0.004 pc from the PDR edge, corresponding to the peak emission in all photometric bands (see the lower panel of Fig.~\ref{fig:best_profiles}). 
The brightness observed at the same position for each JWST filter we are using are also overlaid, together with observed NIRSpec and MIRI/MRS spectra.

The comparison between the modeled and observed spectra shows good agreement across the near-IR spectrum. Given the signal-to-noise of the MIRI/MRS spectrum, we cannot comment on any structure around 5-6 microns. With this limitation, it appears that the overall continuum and bands-to-continuum ratio of the MIRI/MRS spectrum are well reproduced by our model. Additionally, we evaluate the model's fit by comparing the photometric bands from the model with the available observed maximum photometric points values. Consistent with the photometric brightness profile comparison, the model demonstrates general coherence with the observations; however, specific discrepancies are noted, particularly in the 5.6 and 7.7 \(\mu\)m bands, as well as in the continuum around 10 \(\mu\)m and around 25 \(\mu\)m.

The discrepancy between the observed and modeled brightnesses in the F560W filter likely arises from the absence of some emission features between 5 and 6 \(\mu\)m within the THEMIS model (see Fig. 8 of \citealt{Elyajouri2024}). When THEMIS was developed, bands were assigned to various C-C and C-H bonds based on available data \cite{Dartois2004,pino2008,carpentier2012}, and only well-determined features were included \citep{jones2012a,jones2012b,jones2012c}.

At 7.7 \(\mu\)m, the model slightly overestimates the continuum, even as the corresponding brightness profile is underestimated. This discrepancy may stem from the broad spectral coverage of the 7.7 \(\mu\)m filter, which traces the combined emissions from the continuum, the 7.7 \(\mu\)m complex, and 8.6 \(\mu\)m aromatic features. The observed bands in the Horsehead PDR may be broader than those included in the THEMIS model.

A variation of the nano-grains hydrogenation (or bandgap) with size cannot be excluded and might also explain part of the discrepancy. Such variations are not yet included in the THEMIS model.

Around 10 \(\mu\)m, the model underestimates the continuum, where a plateau is expected between 9 and 11 \(\mu\)m. We can note that similar discrepancies at 5.6 and 10 \(\mu\)m were observed in the Orion Bar, as discussed in detail by \cite{Elyajouri2024}. 
The discrepancy around 10 \(\mu\)m band (abundance and shape) could be influenced by the large silicate grains. Adopting \textsc{THEMIS}~2.0 \citep{Ysard2024}, which updates the silicate optical constants, is expected to improve the local fit near \(10~\mu\mathrm{m}\) without altering our constraints on the carbonaceous nano-grain population. We will evaluate this explicitly in future work.

\subsection{The case of a DISM slope \(\alpha = -5\)}
To further our analysis, we specifically test and ultimately rule out the case of the diffuse interstellar medium (DISM) slope, \(\alpha = -5\). For this purpose, we fix \(\alpha\) at -5 and conduct various \chiTOT tests, using all filter bands in some cases, and selectively removing either long-wavelength, short-wavelength, or H\(_2\)-contaminated bands in others. 

Figure~\ref{fig:alpha-5} shows the resulting models in comparison with both photometry and spectroscopy. The results clearly indicate that a DISM slope cannot reproduce all bands simultaneously. Matching the NIR values requires \aminvsg~values comparable to those of the diffuse medium, but this leads to underestimating the MIR emission. Conversely, achieving a good fit in the MIR wavelengths necessitates significantly larger grain sizes for \aminvsg, which fails to reproduce the 3.3 and 3.4 \(\mu\)m features as well as the NIR continuum. This analysis confirms that grain evolution is clear, and the DISM slope \(\alpha = -5\) is not sufficient for this mixture of diffuse grain populations.
\section{Discussion\label{sec:discussion}}
The main results for 
the external filament located 0-5\arcsec\ from the illuminated edge of the Horsehead are the following :

 \begin{enumerate}
 \item  The gas density \(n_0\) is estimated to be  $\sim$ \( n_{0} = 4 \times 10^{5}\,\mathrm{H.cm^{-3}} \)
and \(z_0\) = 0.007 pc at the nano-grain emission peak. 
 \item The nano-grains (i.e. a-C grains) hydrogenation level, $E_\text{g}$, is the same as in the diffuse ISM (0.1 eV).  
\item  The minimum size of the nano-grains, \aminvsg, is comparable (0.35-0.45 nm) to the diffuse ISM value (0.4 nm). 
\item The nano-grains size distribution is less steep ($\alpha$~= -3.5) than in the diffuse ISM ($\alpha$~= -5). 
\item A notable degeneracy exists between the nano-grain dust-to-gas mass ratio (\abvsg) and the PDR line-of-sight depth (\lpdr). This arises because both parameters influence the intensity of near- and mid-infrared emission without altering the spectral shape. Despite this, our analysis constrains \lpdr\, to an upper limit of 0.015 pc, which suggests that nano-grains in this region do not exhibit the depletion seen in other high-UV PDRs. Future observations extending to longer wavelengths could help resolve this degeneracy by better constraining the population of large grains responsible for far-infrared emissions.
 \end{enumerate}

A summary and comparison of the physical parameters for the Horsehead, DISM and the Orion Bar can be found in Table\,\ref{tab:resume2}. 
\subsection{A high-density edge\label{discussion:density}}
The analysis reveals a high density of \( n_{0} = 4 \times 10^{5}\,\mathrm{H.cm^{-3}} \) at the illuminated edge of the Horsehead PDR.
This result aligns with the recent work by \cite{hernandez-vera2023} on ALMA data, which reveals a very sharp transition between molecular and ionized gas, consistent with a high-density environment. Furthermore, we examined whether the density n$_{0}$ = 4 $\times$ 10$^{5}$ H.cm$^{-3}$ obtained inside the PDR is compatible with the emission of the far-infrared fine-structure lines measured by PACS on board Herschel. The ratio of the 158~\mum~[CII] and the 145~\mum~[OI] lines, weakly affected by line opacity and dust attenuation effects along the line of sight, is a good tracer of the local gas density. The beam sizes of the maps obtained in these lines are increasing in order of wavelength 8.8$\arcsec$ and 11$\arcsec$. Thus, Herschel does not allow to resolve the outer filament studied but probes the average gas density inside the PDR. Thus, we use the [CII]158~\mum/[OI] 145~ \mum~ratio to put a constraint on the density by comparing with the predicted values from PDR models \citep{ferland1998,kaufman99,le_petit_model_2006}. For a broad range of densities between 10$^{3}$ and 10$^{6}$ cm$^{-3}$ (typical of PDRs), the [CII]158 \mum/[OI] 145 \mum~ratio varies between 2 and 40 for G$_{0}$ = 100. The observed [CII]158 \mum/[OI] 145 \mum~ratio toward the cut we are considering is about $5\pm2$, compatible with density of a few 10$^{5}$ cm$^{-3}$. There is therefore a good agreement with our result. 

A value of \( z_0 = 0.007 \, \mathrm{pc} \) is required to accurately fit the brightness profiles. This value is approximately nine times smaller than the \( 0.06 \, \mathrm{pc} \) estimated from Spitzer and Herschel observations, a discrepancy likely due to the lower spatial resolution of these instruments \citep{schirmer2020}.

\subsection{Nano-grain hydrogenation}
The analysis of the NIRSpec spectrum from the illuminated edge of the Horsehead reveals that the small grains maintain the same hydrogenation level as those in the diffuse medium ($E_\text{g}$ = 0.1 eV, \(G_0 = 1\)). In contrast, in the atomic region of the Orion Bar, where the radiation field is significantly stronger %(\(G_0 = 10^4\))
(\(G_0 \sim 10^4\)), \cite{Elyajouri2024} observed a lower hydrogen content in the grain structure ($E_\text{g}$ = 0.03 eV). The use of JWST spectroscopic data, with the main constraint being the 3.4/3.3 $\mu$m  band ratio, has clearly demonstrated the efficiency of hydrogen loss in such PDRs. 
The illuminated edge of the Horsehead Nebula, where no significant variation in hydrogen content is observed, suggests that a stronger radiation field, with \(G_0 > 100\), may be required to trigger noticeable dehydrogenation of the small grains.
Observations of NGC 7023 (\(G_0 \sim 1000\)) provide additional evidence of hydrogenation changes in subnanometric carbonaceous particles, such as PAHs. \citet{Pilleri2015} showed that the intensity of the 3.3 $\mu$m band relative to the total PAH emission increases with \(G_0\), while the contribution of the 3.4 $\mu$m band decreases. This trend reflects the progressive loss of hydrogen atoms from the grain structure as the radiation field becomes harder. Similar changes have also been reported in protoplanetary disks \citep{bouteraon2019}, where intense UV radiation plays a role in the dehydrogenation process.

\subsection{Evolution of the nano-grain minimum size and power-law size-distribution exponent\label{discussion:amin}} 

In the Horsehead, our model indicates a %marginally smaller 
similar nano-grain minimum size (\aminvsg~= 0.35-0.45 nm) %(\aminvsg~= 0.35 nm) 
compared to the diffuse ISM \citep[0.4 nm;][]{Jones_2013} and smaller than the one found by \cite{schirmer2020}, who reported 2 to 2.25
times larger values than in the diffuse ISM. 

The context provided by \cite{Abergel2024} is crucial for interpreting these differences. This study shows that the infrared flux of the main bright filaments which are observed behind the illumined edge (Fig.~\ref{fig:HH_maps}) are strongly influenced by the presence of dense matter located behind the illuminated edge of the PDR and the observer (see Fig. 13 of \citealt{Abergel2024}). This dense matter contains dust particles that attenuate the observed brightness in all JWST filters. However, our analysis is focused specifically on the illuminated external edge of the PDR at distances of 0-5\arcsec which presents a  filament fainter than the bright filaments observed at larger distances (5-15\arcsec, see Fig. 1). For this faint filament \cite{Abergel2024} have shown that the dust attenuation is negligible. 

This focus on a region with negligible attenuation is essential for accurately assessing grain sizes and abundances. In contrast, the \cite{schirmer2020} study which used Spitzer and Herschel data was focused on the bright filaments without accounting for dust attenuation. At the time, because the lack of angular resolution and limited spectral coverage and sensitivity, it was not known that the bright infrared filament which were observed were affected by dust attenuation, and as a result, the observed emission was not corrected for this effect. This oversight led to a misinterpretation of the data, where diminished short-wavelength bands—unaccounted for due to extinction—were interpreted as evidence of a truncated distribution with a larger minimum size of $\sim$ 0.8 nm. 

In addition, as shown in Fig.~\ref{fig:s_dist} (left panel), our model indicates a less steep grain size distribution ($\alpha$ = -3.5) compared to the diffuse ISM \citep{Jones_2013} and the Orion bar \citep{Elyajouri2024} where the exponent of the THEMIS nano-grain power-law size distribution is $\alpha$ = -5. 
An increase in $\alpha$, leads to a decrease in the mass of the smallest a-C grains which consequently leads to a decrease in the near-IR emission. Since the total dust mass remains constant, an increase in  $\alpha$ causes a redistribution of the dust mass from the smallest to the largest a-C grains, resulting in an increase in the mid-IR emission. 

The larger value of $\alpha$ found here along with the larger  $E_g$-value shows that the nano-C grains 
are in average larger than in the DISM or the Orion Bar. This suggests that the nano-grains in the Horsehead have been little processed, if at all,  by UV photons or collisions \citep{schirmer2022}.

These results support a model where nano-grain population recovery, potentially through grain reformation due to fragmentation of larger grains, is slower in low and moderate-UV environments, leading to a unique dust size distribution at the edge of the Horsehead Nebula. 
Alternatively, it could be that whatever the process that re-forms nano-grains (the photo- or collisional-fragmentation of larger carbonaceous grains) is less efficient in the Horsehead than in the Orion Bar and in the DISM. In any event, the derived slope of the nano-grain size distribution is intermediate between that of the DISM and that of dense clouds, perhaps signposting the disaggregation of dense cloud dust evolving towards that of the DISM.

\subsection{Evolution of the nano-grain abundance}
We were unable to resolve the degeneracy concerning the abundance of small grains relative to large grains. As illustrated in Fig \ref{fig:spec-all}, this limitation arises because the available JWST filters provide access only to the NIR and MIR wavelength ranges, preventing us from determining the number of large grains. 
This issue could potentially be addressed by observing dust emission at longer wavelengths, as demonstrated by \cite{schirmer2020}. The authors used Herschel/PACS (70 and 160 \mum) and Herschel/SPIRE (250, 350, and 500 \mum) to map the spatial distribution of dust in the Horsehead Nebula across the entire emission spectral range. However, we cannot apply these same filters in our study due to their insufficient resolution to resolve the external isolated filament which spans only 0 to 5 \arcsec. Additionally, as mentioned in the previous section, the overall flux in the region is dominated by the attenuated large filament, rather than the well-exposed small filament that is the focus of our study. Large grains could also be studied through NIR scattering, accessible with HST and JWST filters. However, the available HST and JWST filters are dominated by gas emission lines, which hinders this type of analysis.  

\section{Conclusion\label{sec:conclusion}}
With the data presented in this paper, the behavior of the nano-grains in the Horsehead PDR can be probed both photometrically and spectroscopically. We focussed on the illuminated edge of the Horsehead, an archetypical irradiated PDR. We first examined the NIRSpec spectrum through the 3.3 and 3.4 \mum~ratio and found that the nano-grains are characterized by a band gap energy similar to the typical 0.1 eV in the diffuse ISM.
We studied the Horsehead using 12  photometric bands, from 3.0 to 25.0 \mum, covering the entire NIR and mid-infrared dust spectrum.
Our study presents a detailed model of dust emission at the Horsehead PDR’s illuminated edge, using the THEMIS dust model and SOC radiative transfer code to capture nano-grain properties under moderate UV exposure. Our findings indicate a high-density environment and a less steep size distribution for nano-grains at the illuminated edge, in contrast with the diffuse ISM. This implies that nano-grain destruction mechanisms, such as UV-induced destruction, might be less efficient in the Horsehead's moderate-UV field than in PDRs with more intense radiation, like the Orion Bar. This work provides a template for understanding dust evolution in moderately illuminated PDRs and highlights the need for longer-wavelength observations at high spatial resolutions. Such observations could further constrain the abundance and distribution of nano-grains and larger dust populations, offering insights applicable to other low-UV PDRs.

\begin{acknowledgements}
NIRCam data reduction is performed at the Steward Observatory. MIRI data reduction is performed at the French MIRI centre of expertise with the support of CNES and the ANR-labcom INCLASS between IAS and the company ACRI-ST. This work is based on observations made with the NASA/ESA/CSA \textit{James Webb} Space Telescope. The data were obtained from the Mikulski Archive for Space Telescopes at the Space Telescope Science Institute, which is operated by the Association of Universities for Research in Astronomy, Inc., under NASA contract NAS 5-03127 for JWST. 
Part of this work was supported by the Programme National "Physique et Chimie du Milieu Interstellaire" (PCMI) of CNRS/INSU with INC/INP co-funded by CEA and CNES. 
MEY aknowledges the financial support of CNES. MJ acknowledges the support of the Research Council of Finland grant No. 348342. MB gratefully acknowledges financial support from the Belgian Science Policy Office (BELSPO) through the PRODEX project "JWST/MIRI Science exploitation" (C4000142239).

\end{acknowledgements}

\bibliographystyle{aa}
\bibliography{horsehead}

\begin{thebibliography}{68}
\expandafter\ifx\csname natexlab\endcsname\relax\def\natexlab#1{#1}\fi

\bibitem[{{Abergel} {et~al.}(2010){Abergel}, {Arab}, {Compi{\`e}gne}, {Kirk},
  {Ade}, {Anderson}, {Andr{\'e}}, {Baluteau}, {Bernard}, {Blagrave},
  {Bontemps}, {Boulanger}, {Cohen}, {Cox}, {Dartois}, {Davis}, {Emery},
  {Fulton}, {Gry}, {Habart}, {Huang}, {Joblin}, {Jones}, {Lagache}, {Lim},
  {Madden}, {Makiwa}, {Martin}, {Miville-Desch{\^e}nes}, {Molinari}, {Moseley},
  {Motte}, {Naylor}, {Okumura}, {Pinheiro Gon{\c{c}}alves}, {Polehampton},
  {Rodon}, {Russeil}, {Saraceno}, {Sauvage}, {Sidher}, {Spencer}, {Swinyard},
  {Ward-Thompson}, {White}, \& {Zavagno}}]{abergel2010}
{Abergel}, A., {Arab}, H., {Compi{\`e}gne}, M., {et~al.} 2010, \aap, 518, L96

\bibitem[{{Abergel} {et~al.}(2002){Abergel}, {Bernard}, {Boulanger},
  {Cesarsky}, {Falgarone}, {Jones}, {Miville-Deschenes}, {Perault}, {Puget},
  {Huldtgren}, {Kaas}, {Nordh}, {Olofsson}, {Andr{\'e}}, {Bontemps}, {Casali},
  {Cesarsky}, {Copet}, {Davies}, {Montmerle}, {Persi}, \&
  {Sibille}}]{abergel_isocam_2002}
{Abergel}, A., {Bernard}, J.~P., {Boulanger}, F., {et~al.} 2002, \aap, 389, 239

\bibitem[{{Abergel} {et~al.}(2024){Abergel}, {Misselt}, {Gordon},
  {Noriega-Crespo}, {Guillard}, {Van De Putte}, {Witt}, {Ysard}, {Baes},
  {Beuther}, {Bouchet}, {Brandl}, {Elyajouri}, {Kannavou}, {Kendrew},
  {Klassen}, \& {Trahin}}]{Abergel2024}
{Abergel}, A., {Misselt}, K., {Gordon}, K.~D., {et~al.} 2024, \aap, 687, A4

\bibitem[{{Abergel} {et~al.}(2003){Abergel}, {Teyssier}, {Bernard},
  {Boulanger}, {Coulais}, {Fosse}, {Falgarone}, {Gerin}, {Perault}, {Puget},
  {Nordh}, {Olofsson}, {Huldtgren}, {Kaas}, {Andr{\'e}}, {Bontemps}, {Casali},
  {Cesarsky}, {Copet}, {Davies}, {Montmerle}, {Persi}, \&
  {Sibille}}]{abergel_isocam_2003}
{Abergel}, A., {Teyssier}, D., {Bernard}, J.~P., {et~al.} 2003, \aap, 410, 577

\bibitem[{{Alata} {et~al.}(2014){Alata}, {Cruz-Diaz}, {Mu{\~n}oz Caro}, \&
  {Dartois}}]{Alata2014}
{Alata}, I., {Cruz-Diaz}, G.~A., {Mu{\~n}oz Caro}, G.~M., \& {Dartois}, E.
  2014, \aap, 569, A119

\bibitem[{{Alata} {et~al.}(2015){Alata}, {Jallat}, {Gavilan}, {Chabot},
  {Cruz-Diaz}, {Munoz Caro}, {B{\'e}roff}, \& {Dartois}}]{Alata2015}
{Alata}, I., {Jallat}, A., {Gavilan}, L., {et~al.} 2015, \aap, 584, A123

\bibitem[{Anthony-Twarog(1982)}]{anthony-twarog_h-beta_1982}
Anthony-Twarog, B.~J. 1982, The Astronomical Journal, 87, 1213

\bibitem[{{Arab} {et~al.}(2012){Arab}, {Abergel}, {Habart}, {Bernard-Salas},
  {Ayasso}, {Dassas}, {Martin}, \& {White}}]{Arab2012}
{Arab}, H., {Abergel}, A., {Habart}, E., {et~al.} 2012, \aap, 541, A19

\bibitem[{{Bern{\'e}} {et~al.}(2022){Bern{\'e}}, {Foschino}, {Jalabert}, \&
  {Joblin}}]{Berne2022}
{Bern{\'e}}, O., {Foschino}, S., {Jalabert}, F., \& {Joblin}, C. 2022, \aap,
  667, A159

\bibitem[{{Bern{\'e}} {et~al.}(2007){Bern{\'e}}, {Joblin}, {Deville}, {Smith},
  {Rapacioli}, {Bernard}, {Thomas}, {Reach}, \& {Abergel}}]{berne2007}
{Bern{\'e}}, O., {Joblin}, C., {Deville}, Y., {et~al.} 2007, A\&A, 469, 575

\bibitem[{{Bout{\'e}raon} {et~al.}(2019){Bout{\'e}raon}, {Habart}, {Ysard},
  {Jones}, {Dartois}, \& {Pino}}]{bouteraon2019}
{Bout{\'e}raon}, T., {Habart}, E., {Ysard}, N., {et~al.} 2019, \aap, 623, A135

\bibitem[{{Carpentier} {et~al.}(2012){Carpentier}, {F{\'e}raud}, {Dartois},
  {Brunetto}, {Charon}, {Cao}, {d'Hendecourt}, {Br{\'e}chignac}, {Rouzaud}, \&
  {Pino}}]{carpentier2012}
{Carpentier}, Y., {F{\'e}raud}, G., {Dartois}, E., {et~al.} 2012, \aap, 548,
  A40

\bibitem[{{Compi{\`e}gne} {et~al.}(2008){Compi{\`e}gne}, {Abergel},
  {Verstraete}, \& {Habart}}]{compiegne2008}
{Compi{\`e}gne}, M., {Abergel}, A., {Verstraete}, L., \& {Habart}, E. 2008,
  \aap, 491, 797

\bibitem[{Compi{\`e}gne {et~al.}(2007)Compi{\`e}gne, Abergel, Verstraete,
  Reach, Habart, Smith, Boulanger, \& Joblin}]{compiegne_aromatic_2007}
Compi{\`e}gne, M., Abergel, A., Verstraete, L., {et~al.} 2007, \aap, 471, 205

\bibitem[{{Compi{\`e}gne} {et~al.}(2011){Compi{\`e}gne}, {Verstraete}, {Jones},
  {Bernard}, {Boulanger}, {Flagey}, {Le Bourlot}, {Paradis}, \&
  {Ysard}}]{compiegne2011}
{Compi{\`e}gne}, M., {Verstraete}, L., {Jones}, A., {et~al.} 2011, \aap, 525,
  A103

\bibitem[{{Dartois} {et~al.}(2004){Dartois}, {Mu{\~n}oz Caro}, {Deboffle}, \&
  {d'Hendecourt}}]{Dartois2004}
{Dartois}, E., {Mu{\~n}oz Caro}, G.~M., {Deboffle}, D., \& {d'Hendecourt}, L.
  2004, \aap, 423, L33

\bibitem[{{Draine}(1978)}]{Draine1978}
{Draine}, B.~T. 1978, \apjs, 36, 595

\bibitem[{Duley {et~al.}(2015)Duley, Zaidi, Wesolowski, \& Kuzmin}]{Duley2015}
Duley, W.~W., Zaidi, A., Wesolowski, M.~J., \& Kuzmin, S. 2015, Monthly Notices
  of the Royal Astronomical Society, 447, 1242

\bibitem[{{Elyajouri} {et~al.}(2024){Elyajouri}, {Ysard}, {Abergel}, {Habart},
  {Verstraete}, {Jones}, {Juvela}, {Schirmer}, {Meshaka}, {Dartois},
  {Lebourlot}, {Rouill{\'e}}, {Onaka}, {Peeters}, {Bern{\'e}}, {Alarc{\'o}n},
  {Bernard-Salas}, {Buragohain}, {Cami}, {Canin}, {Chown}, {Demyk}, {Gordon},
  {Kannavou}, {Kirsanova}, {Madden}, {Paladini}, {Pendleton}, {Salama},
  {Schroetter}, {Sidhu}, {R{\"o}llig}, {Trahin}, \& {Van De
  Putte}}]{Elyajouri2024}
{Elyajouri}, M., {Ysard}, N., {Abergel}, A., {et~al.} 2024, \aap, 685, A76

\bibitem[{{Ferland} {et~al.}(1998){Ferland}, {Korista}, {Verner}, {Ferguson},
  {Kingdon}, \& {Verner}}]{ferland1998}
{Ferland}, G.~J., {Korista}, K.~T., {Verner}, D.~A., {et~al.} 1998, \pasp, 110,
  761

\bibitem[{{Gaia Collaboration} {et~al.}(2018){Gaia Collaboration}, {Brown},
  {Vallenari}, {Prusti}, {de Bruijne}, {Babusiaux}, {Bailer-Jones}, {Biermann},
  {Evans}, {Eyer}, {Jansen}, {Jordi}, {Klioner}, {Lammers}, {Lindegren},
  {Luri}, {Mignard}, {Panem}, {Pourbaix}, {Randich}, {Sartoretti}, {Siddiqui},
  {Soubiran}, {van Leeuwen}, {Walton}, {Arenou}, {Bastian}, {Cropper},
  {Drimmel}, {Katz}, {Lattanzi}, {Bakker}, {Cacciari}, {Casta{\~n}eda},
  {Chaoul}, {Cheek}, {De Angeli}, {Fabricius}, {Guerra}, {Holl}, {Masana},
  {Messineo}, {Mowlavi}, {Nienartowicz}, {Panuzzo}, {Portell}, {Riello},
  {Seabroke}, {Tanga}, {Th{\'e}venin}, {Gracia-Abril}, {Comoretto},
  {Garcia-Reinaldos}, {Teyssier}, {Altmann}, {Andrae}, {Audard},
  {Bellas-Velidis}, {Benson}, {Berthier}, {Blomme}, {Burgess}, {Busso},
  {Carry}, {Cellino}, {Clementini}, {Clotet}, {Creevey}, {Davidson}, {De
  Ridder}, {Delchambre}, {Dell'Oro}, {Ducourant},
  {Fern{\'a}ndez-Hern{\'a}ndez}, {Fouesneau}, {Fr{\'e}mat}, {Galluccio},
  {Garc{\'\i}a-Torres}, {Gonz{\'a}lez-N{\'u}{\~n}ez}, {Gonz{\'a}lez-Vidal},
  {Gosset}, {Guy}, {Halbwachs}, {Hambly}, {Harrison}, {Hern{\'a}ndez},
  {Hestroffer}, {Hodgkin}, {Hutton}, {Jasniewicz}, {Jean-Antoine-Piccolo},
  {Jordan}, {Korn}, {Krone-Martins}, {Lanzafame}, {Lebzelter}, {L{\"o}ffler},
  {Manteiga}, {Marrese}, {Mart{\'\i}n-Fleitas}, {Moitinho}, {Mora}, {Muinonen},
  {Osinde}, {Pancino}, {Pauwels}, {Petit}, {Recio-Blanco}, {Richards},
  {Rimoldini}, {Robin}, {Sarro}, {Siopis}, {Smith}, {Sozzetti}, {S{\"u}veges},
  {Torra}, {van Reeven}, {Abbas}, {Abreu Aramburu}, {Accart}, {Aerts},
  {Altavilla}, {{\'A}lvarez}, {Alvarez}, {Alves}, {Anderson}, {Andrei},
  {Anglada Varela}, {Antiche}, {Antoja}, {Arcay}, {Astraatmadja}, {Bach},
  {Baker}, {Balaguer-N{\'u}{\~n}ez}, {Balm}, {Barache}, {Barata}, {Barbato},
  {Barblan}, {Barklem}, {Barrado}, {Barros}, {Barstow}, {Bartholom{\'e}
  Mu{\~n}oz}, {Bassilana}, {Becciani}, {Bellazzini}, {Berihuete}, {Bertone},
  {Bianchi}, {Bienaym{\'e}}, {Blanco-Cuaresma}, {Boch}, {Boeche}, {Bombrun},
  {Borrachero}, {Bossini}, {Bouquillon}, {Bourda}, {Bragaglia}, {Bramante},
  {Breddels}, {Bressan}, {Brouillet}, {Br{\"u}semeister}, {Brugaletta},
  {Bucciarelli}, {Burlacu}, {Busonero}, {Butkevich}, {Buzzi}, {Caffau},
  {Cancelliere}, {Cannizzaro}, {Cantat-Gaudin}, {Carballo}, {Carlucci},
  {Carrasco}, {Casamiquela}, {Castellani}, {Castro-Ginard}, {Charlot},
  {Chemin}, {Chiavassa}, {Cocozza}, {Costigan}, {Cowell}, {Crifo}, {Crosta},
  {Crowley}, {Cuypers}, {Dafonte}, {Damerdji}, {Dapergolas}, {David}, {David},
  {de Laverny}, {De Luise}, {De March}, {de Martino}, {de Souza}, {de Torres},
  {Debosscher}, {del Pozo}, {Delbo}, {Delgado}, {Delgado}, {Di Matteo},
  {Diakite}, {Diener}, {Distefano}, {Dolding}, {Drazinos}, {Dur{\'a}n},
  {Edvardsson}, {Enke}, {Eriksson}, {Esquej}, {Eynard Bontemps}, {Fabre},
  {Fabrizio}, {Faigler}, {Falc{\~a}o}, {Farr{\`a}s Casas}, {Federici},
  {Fedorets}, {Fernique}, {Figueras}, {Filippi}, {Findeisen}, {Fonti},
  {Fraile}, {Fraser}, {Fr{\'e}zouls}, {Gai}, {Galleti}, {Garabato},
  {Garc{\'\i}a-Sedano}, {Garofalo}, {Garralda}, {Gavel}, {Gavras}, {Gerssen},
  {Geyer}, {Giacobbe}, {Gilmore}, {Girona}, {Giuffrida}, {Glass}, {Gomes},
  {Granvik}, {Gueguen}, {Guerrier}, {Guiraud}, {Guti{\'e}rrez-S{\'a}nchez},
  {Haigron}, {Hatzidimitriou}, {Hauser}, {Haywood}, {Heiter}, {Helmi}, {Heu},
  {Hilger}, {Hobbs}, {Hofmann}, {Holland}, {Huckle}, {Hypki}, {Icardi},
  {Jan{\ss}en}, {Jevardat de Fombelle}, {Jonker}, {Juh{\'a}sz}, {Julbe},
  {Karampelas}, {Kewley}, {Klar}, {Kochoska}, {Kohley}, {Kolenberg},
  {Kontizas}, {Kontizas}, {Koposov}, {Kordopatis}, {Kostrzewa-Rutkowska},
  {Koubsky}, {Lambert}, {Lanza}, {Lasne}, {Lavigne}, {Le Fustec}, {Le
  Poncin-Lafitte}, {Lebreton}, {Leccia}, {Leclerc}, {Lecoeur-Taibi},
  {Lenhardt}, {Leroux}, {Liao}, {Licata}, {Lindstr{\o}m}, {Lister}, {Livanou},
  {Lobel}, {L{\'o}pez}, {Managau}, {Mann}, {Mantelet}, {Marchal}, {Marchant},
  {Marconi}, {Marinoni}, {Marschalk{\'o}}, {Marshall}, {Martino}, {Marton},
  {Mary}, {Massari}, {Matijevi{\v{c}}}, {Mazeh}, {McMillan}, {Messina},
  {Michalik}, {Millar}, {Molina}, {Molinaro}, {Moln{\'a}r}, {Montegriffo},
  {Mor}, {Morbidelli}, {Morel}, {Morris}, {Mulone}, {Muraveva}, {Musella},
  {Nelemans}, {Nicastro}, {Noval}, {O'Mullane}, {Ord{\'e}novic},
  {Ord{\'o}{\~n}ez-Blanco}, {Osborne}, {Pagani}, {Pagano}, {Pailler},
  {Palacin}, {Palaversa}, {Panahi}, {Pawlak}, {Piersimoni}, {Pineau}, {Plachy},
  {Plum}, {Poggio}, {Poujoulet}, {Pr{\v{s}}a}, {Pulone}, {Racero}, {Ragaini},
  {Rambaux}, {Ramos-Lerate}, {Regibo}, {Reyl{\'e}}, {Riclet}, {Ripepi}, {Riva},
  {Rivard}, {Rixon}, {Roegiers}, {Roelens}, {Romero-G{\'o}mez}, {Rowell},
  {Royer}, {Ruiz-Dern}, {Sadowski}, {Sagrist{\`a} Sell{\'e}s}, {Sahlmann},
  {Salgado}, {Salguero}, {Sanna}, {Santana-Ros}, {Sarasso}, {Savietto},
  {Schultheis}, {Sciacca}, {Segol}, {Segovia}, {S{\'e}gransan}, {Shih},
  {Siltala}, {Silva}, {Smart}, {Smith}, {Solano}, {Solitro}, {Sordo}, {Soria
  Nieto}, {Souchay}, {Spagna}, {Spoto}, {Stampa}, {Steele},
  {Steidelm{\"u}ller}, {Stephenson}, {Stoev}, {Suess}, {Surdej}, {Szabados},
  {Szegedi-Elek}, {Tapiador}, {Taris}, {Tauran}, {Taylor}, {Teixeira},
  {Terrett}, {Teyssandier}, {Thuillot}, {Titarenko}, {Torra Clotet}, {Turon},
  {Ulla}, {Utrilla}, {Uzzi}, {Vaillant}, {Valentini}, {Valette}, {van Elteren},
  {Van Hemelryck}, {van Leeuwen}, {Vaschetto}, {Vecchiato}, {Veljanoski},
  {Viala}, {Vicente}, {Vogt}, {von Essen}, {Voss}, {Votruba}, {Voutsinas},
  {Walmsley}, {Weiler}, {Wertz}, {Wevers}, {Wyrzykowski}, {Yoldas},
  {{\v{Z}}erjal}, {Ziaeepour}, {Zorec}, {Zschocke}, {Zucker}, {Zurbach}, \&
  {Zwitter}}]{2018A&A...616A...1G}
{Gaia Collaboration}, {Brown}, A.~G.~A., {Vallenari}, A., {et~al.} 2018, \aap,
  616, A1

\bibitem[{{Galliano} {et~al.}(2008){Galliano}, {Madden}, {Tielens}, {Peeters},
  \& {Jones}}]{Galliano:08}
{Galliano}, F., {Madden}, S.~C., {Tielens}, A. G.~G.~M., {Peeters}, E., \&
  {Jones}, A.~P. 2008, \apj, 679, 310

\bibitem[{{Gerin} {et~al.}(2009){Gerin}, {Goicoechea}, {Pety}, \&
  {Hily-Blant}}]{Gerin2009}
{Gerin}, M., {Goicoechea}, J.~R., {Pety}, J., \& {Hily-Blant}, P. 2009, \aap,
  494, 977

\bibitem[{{Goicoechea} {et~al.}(2016){Goicoechea}, {Pety}, {Cuadrado},
  {Cernicharo}, {Chapillon}, {Fuente}, {Gerin}, {Joblin}, {Marcelino}, \&
  {Pilleri}}]{2016Natur.537..207G}
{Goicoechea}, J.~R., {Pety}, J., {Cuadrado}, S., {et~al.} 2016, \nat, 537, 207

\bibitem[{{Goicoechea} {et~al.}(2006){Goicoechea}, {Pety}, {Gerin}, {Teyssier},
  {Roueff}, {Hily-Blant}, \& {Baek}}]{goicoechea2006}
{Goicoechea}, J.~R., {Pety}, J., {Gerin}, M., {et~al.} 2006, \aap, 456, 565

\bibitem[{{Gordon} {et~al.}(2025){Gordon}, {Sloan}, {Garcia Marin},
  {Libralato}, {Rieke}, {Aguilar}, {Bohlin}, {Cracraft}, {Decleir}, {Gaspar},
  {Kendrew}, {Law}, {Noriega-Crespo}, \& {Regan}}]{Gordon2025}
{Gordon}, K.~D., {Sloan}, G.~C., {Garcia Marin}, M., {et~al.} 2025, \aj, 169, 6

\bibitem[{Guzmán {et~al.}(2011)Guzmán, Pety, Goicoechea, Gerin, \&
  Roueff}]{Guzman2011}
Guzmán, V., Pety, J., Goicoechea, J.~R., Gerin, M., \& Roueff, E. 2011,
  Astronomy and Astrophysics, 534, A49

\bibitem[{{Habart} {et~al.}(2005{\natexlab{a}}){Habart}, {Abergel}, {Walmsley},
  {Teyssier}, \& {Pety}}]{habart_2005}
{Habart}, E., {Abergel}, A., {Walmsley}, C.~M., {Teyssier}, D., \& {Pety}, J.
  2005{\natexlab{a}}, \aap, 437, 177

\bibitem[{{Habart} {et~al.}(2023){Habart}, {Peeters}, {Bern{\'e}}, {Trahin},
  {Canin}, {Chown}, {Sidhu}, {Van De Putte}, {Alarc{\'o}n}, {Schroetter},
  {Dartois}, {Vicente}, {Abergel}, {Bergin}, {Bernard-Salas}, {Boersma},
  {Bron}, {Cami}, {Cuadrado}, {Dicken}, {Elyajouri}, {Fuente}, {Goicoechea},
  {Gordon}, {Issa}, {Joblin}, {Kannavou}, {Khan}, {Lacinbala}, {Languignon},
  {Le Gal}, {Maragkoudakis}, {Meshaka}, {Okada}, {Onaka}, {Pasquini}, {Pound},
  {Robberto}, {R{\"o}llig}, {Schefter}, {Schirmer}, {Tabone}, {Tielens},
  {Wolfire}, {Zannese}, {Ysard}, {Miville-Deschenes}, {Aleman}, {Allamandola},
  {Auchettl}, {Baratta}, {Bejaoui}, {Bera}, {Black}, {Boulanger}, {Bouwman},
  {Brandl}, {Brechignac}, {Br{\"u}nken}, {Buragohain}, {Burkhardt}, {Candian},
  {Cazaux}, {Cernicharo}, {Chabot}, {Chakraborty}, {Champion}, {Colgan},
  {Cooke}, {Coutens}, {Cox}, {Demyk}, {Donovan Meyer}, {Foschino},
  {Garc{\'\i}a-Lario}, {Gavilan}, {Gerin}, {Gottlieb}, {Guillard}, {Gusdorf},
  {Hartigan}, {He}, {Herbst}, {Hornekaer}, {J{\"a}ger}, {Janot-Pacheco},
  {Kaufman}, {Kemper}, {Kendrew}, {Kirsanova}, {Klaassen}, {Kwok}, {Labiano},
  {Lai}, {Lee}, {Lefloch}, {Le Petit}, {Li}, {Linz}, {Mackie}, {Madden},
  {Mascetti}, {McGuire}, {Merino}, {Micelotta}, {Misselt}, {Morse}, {Mulas},
  {Neelamkodan}, {Ohsawa}, {Omont}, {Paladini}, {Palumbo}, {Pathak},
  {Pendleton}, {Petrignani}, {Pino}, {Puga}, {Rangwala}, {Rapacioli}, {Ricca},
  {Roman-Duval}, {Roser}, {Roueff}, {Rouill{\'e}}, {Salama}, {Sales},
  {Sandstrom}, {Sarre}, {Sciamma-O'Brien}, {Sellgren}, {Shenoy}, {Teyssier},
  {Thomas}, {Togi}, {Verstraete}, {Witt}, {Wootten}, {Zettergren}, {Zhang},
  {Zhang}, \& {Zhen}}]{Habart2023jwst}
{Habart}, E., {Peeters}, E., {Bern{\'e}}, O., {et~al.} 2023, arXiv e-prints,
  arXiv:2308.16732

\bibitem[{{Habart} {et~al.}(2005{\natexlab{b}}){Habart}, {Walmsley},
  {Verstraete}, {Cazaux}, {Maiolino}, {Cox}, {Boulanger}, \& {Pineau des
  For{\^e}ts}}]{habart05}
{Habart}, E., {Walmsley}, M., {Verstraete}, L., {et~al.} 2005{\natexlab{b}},
  \ssr, 119, 71

\bibitem[{{Hern{\'a}ndez-Vera} {et~al.}(2023){Hern{\'a}ndez-Vera},
  {Guzm{\'a}n}, {Goicoechea}, {Maillard}, {Pety}, {Le Petit}, {Gerin}, {Bron},
  {Roueff}, {Abergel}, {Schirmer}, {Carpenter}, {Gratier}, {Gordon}, \&
  {Misselt}}]{hernandez-vera2023}
{Hern{\'a}ndez-Vera}, C., {Guzm{\'a}n}, V.~V., {Goicoechea}, J.~R., {et~al.}
  2023, \aap, 677, A152

\bibitem[{{Hollenbach} \& {Tielens}(1997)}]{hollenbach1997a}
{Hollenbach}, D.~J. \& {Tielens}, A.~G.~G.~M. 1997, \araa, 35, 179

\bibitem[{Hollenbach \& Tielens(1999)}]{hollenbach1999}
Hollenbach, D.~J. \& Tielens, A. G. G.~M. 1999, Reviews of Modern Physics, 71,
  173

\bibitem[{{Joblin} {et~al.}(1996){Joblin}, {Tielens}, {Geballe}, \&
  {Wooden}}]{Joblin96_ngc1333}
{Joblin}, C., {Tielens}, A.~G.~G.~M., {Geballe}, T.~R., \& {Wooden}, D.~H.
  1996, \apjl, 460, L119

\bibitem[{{Jones}(2012{\natexlab{a}})}]{jones2012a}
{Jones}, A.~P. 2012{\natexlab{a}}, \aap, 540, A1

\bibitem[{{Jones}(2012{\natexlab{b}})}]{jones2012b}
{Jones}, A.~P. 2012{\natexlab{b}}, \aap, 540, A2

\bibitem[{{Jones}(2012{\natexlab{c}})}]{jones2012c}
{Jones}, A.~P. 2012{\natexlab{c}}, \aap, 542, A98

\bibitem[{{Jones}(2014)}]{jones2014}
{Jones}, A.~P. 2014, \planss, 100, 26

\bibitem[{{Jones} {et~al.}(2013){Jones}, {Fanciullo}, {K{\"o}hler},
  {Verstraete}, {Guillet}, {Bocchio}, \& {Ysard}}]{Jones_2013}
{Jones}, A.~P., {Fanciullo}, L., {K{\"o}hler}, M., {et~al.} 2013, \aap, 558,
  A62

\bibitem[{{Jones} {et~al.}(2017){Jones}, {K{\"o}hler}, {Ysard}, {Bocchio}, \&
  {Verstraete}}]{jones2017}
{Jones}, A.~P., {K{\"o}hler}, M., {Ysard}, N., {Bocchio}, M., \& {Verstraete},
  L. 2017, \aap, 602, A46

\bibitem[{{Juvela}(2019)}]{juvela2019}
{Juvela}, M. 2019, \aap, 622, A79

\bibitem[{{Kaufman} {et~al.}(1999){Kaufman}, {Wolfire}, {Hollenbach}, \&
  {Luhman}}]{kaufman99}
{Kaufman}, M.~J., {Wolfire}, M.~G., {Hollenbach}, D.~J., \& {Luhman}, M.~L.
  1999, \apj, 527, 795

\bibitem[{{K{\"o}hler} {et~al.}(2015){K{\"o}hler}, {Ysard}, \&
  {Jones}}]{kohler15}
{K{\"o}hler}, M., {Ysard}, N., \& {Jones}, A.~P. 2015, \aap, 579, A15

\bibitem[{{Le Gal} {et~al.}(2017){Le Gal}, {Herbst}, {Dufour}, {Gratier},
  {Ruaud}, {Vidal}, \& {Wakelam}}]{LeGal2017}
{Le Gal}, R., {Herbst}, E., {Dufour}, G., {et~al.} 2017, \aap, 605, A88

\bibitem[{Le~Petit {et~al.}(2006)Le~Petit, Nehme, Le~Bourlot, \&
  Roueff}]{le_petit_model_2006}
Le~Petit, F., Nehme, C., Le~Bourlot, J., \& Roueff, E. 2006, The Astrophysical
  Journal Supplement Series, 164, 506

\bibitem[{{Marconi} {et~al.}(1998){Marconi}, {Testi}, {Natta}, \&
  {Walmsley}}]{Marconi1998}
{Marconi}, A., {Testi}, L., {Natta}, A., \& {Walmsley}, C.~M. 1998, \aap, 330,
  696

\bibitem[{{Misselt} {et~al.}(2025){Misselt}, {Witt}, {Gordon}, {Van De Putte},
  {Trahin}, {Abergel}, {Noriega-Crespo}, {Guillard}, {Zannese}, {Dell'ova},
  {Baes}, {Klaassen}, \& {Ysard}}]{Misselt2025}
{Misselt}, K., {Witt}, A.~N., {Gordon}, K.~D., {et~al.} 2025, \aap, 700, A158

\bibitem[{Ohashi {et~al.}(2013)Ohashi, Kitamura, \& Akashi}]{Ohashi2013}
Ohashi, S., Kitamura, Y., \& Akashi, T. 2013, in , 345

\bibitem[{{Peeters} {et~al.}(2004){Peeters}, {Allamandola}, {Bauschlicher},
  {Hudgins}, {Sandford}, \& {Tielens}}]{Peeters04}
{Peeters}, E., {Allamandola}, L.~J., {Bauschlicher}, C.~W., J., {et~al.} 2004,
  \apj, 604, 252

\bibitem[{{Peeters} {et~al.}(2024){Peeters}, {Habart}, {Bern{\'e}}, {Sidhu},
  {Chown}, {Van De Putte}, {Trahin}, {Schroetter}, {Canin}, {Alarc{\'o}n},
  {Schefter}, {Khan}, {Pasquini}, {Tielens}, {Wolfire}, {Dartois},
  {Goicoechea}, {Maragkoudakis}, {Onaka}, {Pound}, {Vicente}, {Abergel},
  {Bergin}, {Bernard-Salas}, {Boersma}, {Bron}, {Cami}, {Cuadrado}, {Dicken},
  {Elyajouri}, {Fuente}, {Gordon}, {Issa}, {Joblin}, {Kannavou}, {Lacinbala},
  {Languignon}, {Le Gal}, {Meshaka}, {Okada}, {Robberto}, {R{\"o}llig},
  {Schirmer}, {Tabone}, {Zannese}, {Aleman}, {Allamandola}, {Auchettl},
  {Baratta}, {Bejaoui}, {Bera}, {Black}, {Boulanger}, {Bouwman}, {Brandl},
  {Brechignac}, {Br{\"u}nken}, {Buragohain}, {Burkhardt}, {Candian}, {Cazaux},
  {Cernicharo}, {Chabot}, {Chakraborty}, {Champion}, {Colgan}, {Cooke},
  {Coutens}, {Cox}, {Demyk}, {Meyer}, {Foschino}, {Garc{\'\i}a-Lario}, {Gerin},
  {Gottlieb}, {Guillard}, {Gusdorf}, {Hartigan}, {He}, {Herbst}, {Hornekaer},
  {J{\"a}ger}, {Janot-Pacheco}, {Kaufman}, {Kendrew}, {Kirsanova}, {Klaassen},
  {Kwok}, {Labiano}, {Lai}, {Lee}, {Lefloch}, {Le Petit}, {Li}, {Linz},
  {Mackie}, {Madden}, {Mascetti}, {McGuire}, {Merino}, {Micelotta}, {Misselt},
  {Morse}, {Mulas}, {Neelamkodan}, {Ohsawa}, {Paladini}, {Palumbo}, {Pathak},
  {Pendleton}, {Petrignani}, {Pino}, {Puga}, {Rangwala}, {Rapacioli}, {Ricca},
  {Roman-Duval}, {Roser}, {Roueff}, {Rouill{\'e}}, {Salama}, {Sales},
  {Sandstrom}, {Sarre}, {Sciamma-O'Brien}, {Sellgren}, {Shenoy}, {Teyssier},
  {Thomas}, {Togi}, {Verstraete}, {Witt}, {Wootten}, {Ysard}, {Zettergren},
  {Zhang}, {Zhang}, \& {Zhen}}]{peeters2024}
{Peeters}, E., {Habart}, E., {Bern{\'e}}, O., {et~al.} 2024, \aap, 685, A74

\bibitem[{Peeters {et~al.}(2002)Peeters, Hony, Van~Kerckhoven, Tielens,
  Allamandola, Hudgins, \& Bauschlicher}]{peeters_rich_2002}
Peeters, E., Hony, S., Van~Kerckhoven, C., {et~al.} 2002, \aap, 390, 1089

\bibitem[{Perrin {et~al.}(2014)Perrin, Sivaramakrishnan, Lajoie, Elliott,
  Pueyo, Ravindranath, \& Albert}]{Perrin20214}
Perrin, M.~D., Sivaramakrishnan, A., Lajoie, C.-P., {et~al.} 2014, in Space
  Telescopes and Instrumentation 2014: Optical, Infrared, and Millimeter Wave,
  ed. J.~M.~O. Jr., M.~Clampin, G.~G. Fazio, \& H.~A. MacEwen, Vol. 9143,
  International Society for Optics and Photonics (SPIE), 91433X

\bibitem[{{Pety} {et~al.}(2012){Pety}, {Gratier}, {Guzm{\'a}n}, {Roueff},
  {Gerin}, {Goicoechea}, {Bardeau}, {Sievers}, {Le Petit}, {Le Bourlot},
  {Belloche}, \& {Talbi}}]{Pety2012}
{Pety}, J., {Gratier}, P., {Guzm{\'a}n}, V., {et~al.} 2012, \aap, 548, A68

\bibitem[{Pety {et~al.}(2005)Pety, Teyssier, Fossé, Gerin, Roueff, Abergel,
  Habart, \& Cernicharo}]{pety_are_2005}
Pety, J., Teyssier, D., Fossé, D., {et~al.} 2005, Astronomy \& Astrophysics,
  435, 885

\bibitem[{{Pilleri} {et~al.}(2015){Pilleri}, {Joblin}, {Boulanger}, \&
  {Onaka}}]{Pilleri2015}
{Pilleri}, P., {Joblin}, C., {Boulanger}, F., \& {Onaka}, T. 2015, \aap, 577,
  A16

\bibitem[{{Pino} {et~al.}(2008){Pino}, {Dartois}, {Cao}, {Carpentier},
  {Chamaill{\'e}}, {Vasquez}, {Jones}, {D'Hendecourt}, \&
  {Br{\'e}chignac}}]{pino2008}
{Pino}, T., {Dartois}, E., {Cao}, A.~T., {et~al.} 2008, \aap, 490, 665

\bibitem[{{Rapacioli} {et~al.}(2006){Rapacioli}, {Calvo}, {Joblin}, {Parneix},
  {Toublanc}, \& {Spiegelman}}]{Rapacioli2006}
{Rapacioli}, M., {Calvo}, F., {Joblin}, C., {et~al.} 2006, \aap, 460, 519

\bibitem[{Schaerer \& de~Koter(1997)}]{schaerer_combined_1997}
Schaerer, D. \& de~Koter, A. 1997, Astronomy and Astrophysics, 322

\bibitem[{Schirmer {et~al.}(2020)Schirmer, Abergel, Verstraete, Ysard, Juvela,
  Jones, \& Habart}]{schirmer2020}
Schirmer, T., Abergel, A., Verstraete, L., {et~al.} 2020, Astronomy \&
  Astrophysics, 639, A144

\bibitem[{{Schirmer} {et~al.}(2020){Schirmer}, {Abergel}, {Verstraete},
  {Ysard}, {Juvela}, {Jones}, \& {Habart}}]{schirmer_2020}
{Schirmer}, T., {Abergel}, A., {Verstraete}, L., {et~al.} 2020, \aap, 639, A144

\bibitem[{{Schirmer} {et~al.}(2022){Schirmer}, {Ysard}, {Habart}, {Jones},
  {Abergel}, \& {Verstraete}}]{schirmer2022}
{Schirmer}, T., {Ysard}, N., {Habart}, E., {et~al.} 2022, \aap, 666, A49

\bibitem[{{Smith}(1984)}]{smith1984}
{Smith}, F.~W. 1984, Journal of Applied Physics, 55, 764

\bibitem[{{Teyssier} {et~al.}(2004){Teyssier}, {Foss{\'e}}, {Gerin}, {Pety},
  {Abergel}, \& {Roueff}}]{Teyssier2004}
{Teyssier}, D., {Foss{\'e}}, D., {Gerin}, M., {et~al.} 2004, \aap, 417, 135

\bibitem[{Warren \& Hesser(1977)}]{warren_photometric_1977}
Warren, Jr., W.~H. \& Hesser, J.~E. 1977, The Astrophysical Journal Supplement
  Series, 34

\bibitem[{{Wolfire} {et~al.}(2022){Wolfire}, {Vallini}, \&
  {Chevance}}]{wolfire2022}
{Wolfire}, M.~G., {Vallini}, L., \& {Chevance}, M. 2022, \araa, 60, 247

\bibitem[{{Ysard} {et~al.}(2024){Ysard}, {Jones}, {Guillet}, {Demyk},
  {Decleir}, {Verstraete}, {Choubani}, {Miville-Desch{\^e}nes}, \&
  {Fanciullo}}]{Ysard2024}
{Ysard}, N., {Jones}, A.~P., {Guillet}, V., {et~al.} 2024, \aap, 684, A34

\bibitem[{{Ysard} {et~al.}(2015){Ysard}, {K{\"o}hler}, {Jones},
  {Miville-Desch{\^e}nes}, {Abergel}, \& {Fanciullo}}]{Ysard2015}
{Ysard}, N., {K{\"o}hler}, M., {Jones}, A., {et~al.} 2015, \aap, 577, A110

\bibitem[{{Zubko} {et~al.}(2004){Zubko}, {Dwek}, \& {Arendt}}]{Zubko2004}
{Zubko}, V., {Dwek}, E., \& {Arendt}, R.~G. 2004, \apjs, 152, 211

\end{thebibliography}

\begin{appendix}
\section{The contribution of H$_2$ lines in the different NIRCam and MIRI filters}
\begin{table*}[h]
 \caption{Average contribution of H$_2$ lines for each JWST filter in the external filament located 0-5\arcsec\ from the illuminated edge of the Horsehead. Line contribution percentages were derived from \cite{Misselt2025}.}\label{tab:H2_contamination}
    \centering
    
    \begin{tabular}{l c r}
        \hline
        Filter & Wavelength (\mum) & H$_2$ Contribution \\
        \hline
        F300M  & 3.0   & 0.51\\
        F335M  & 3.35  & 0.095 \\
        F430M & 4.3   & 0.212\\
        F560W  & 5.6   & 0.06\\
        F770W  & 7.7   & 0.067\\
        F1000W & 10.0 & 0.19 \\
        F1130W & 11.3  & 0.00 \\
        F1280W & 12.8  & 0.09\\
        F1500W & 15.0  & 0.005 \\
        F1800W & 18.0 & 0.087 \\
        F2100W & 21.0  & 0.00\\
        F2550W & 25.5  & 0.00 \\
        \hline
    \end{tabular}

\end{table*}

\section{FWHM for each band as a function of nano-grains properties (\abvsg, \aminvsg, $\alpha$) and density profile (\(n_{0}\),\(z_{0}\)}
\begin{figure*}
    \centering
    \includegraphics[width=\textwidth]{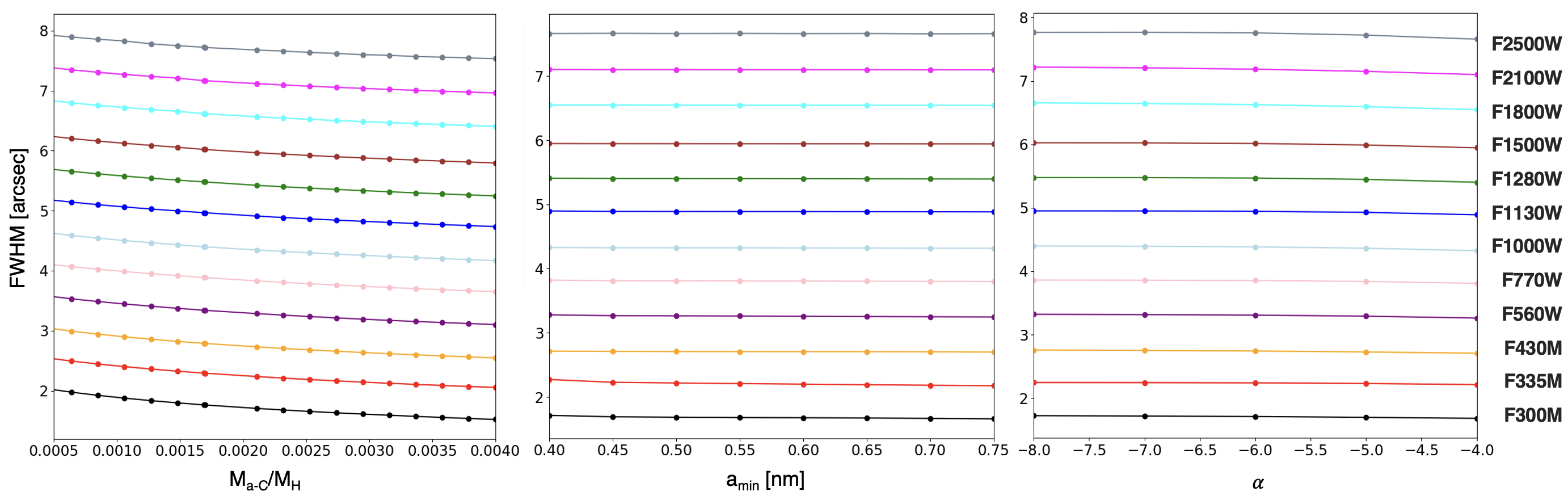}
    \includegraphics[width=0.45\linewidth]{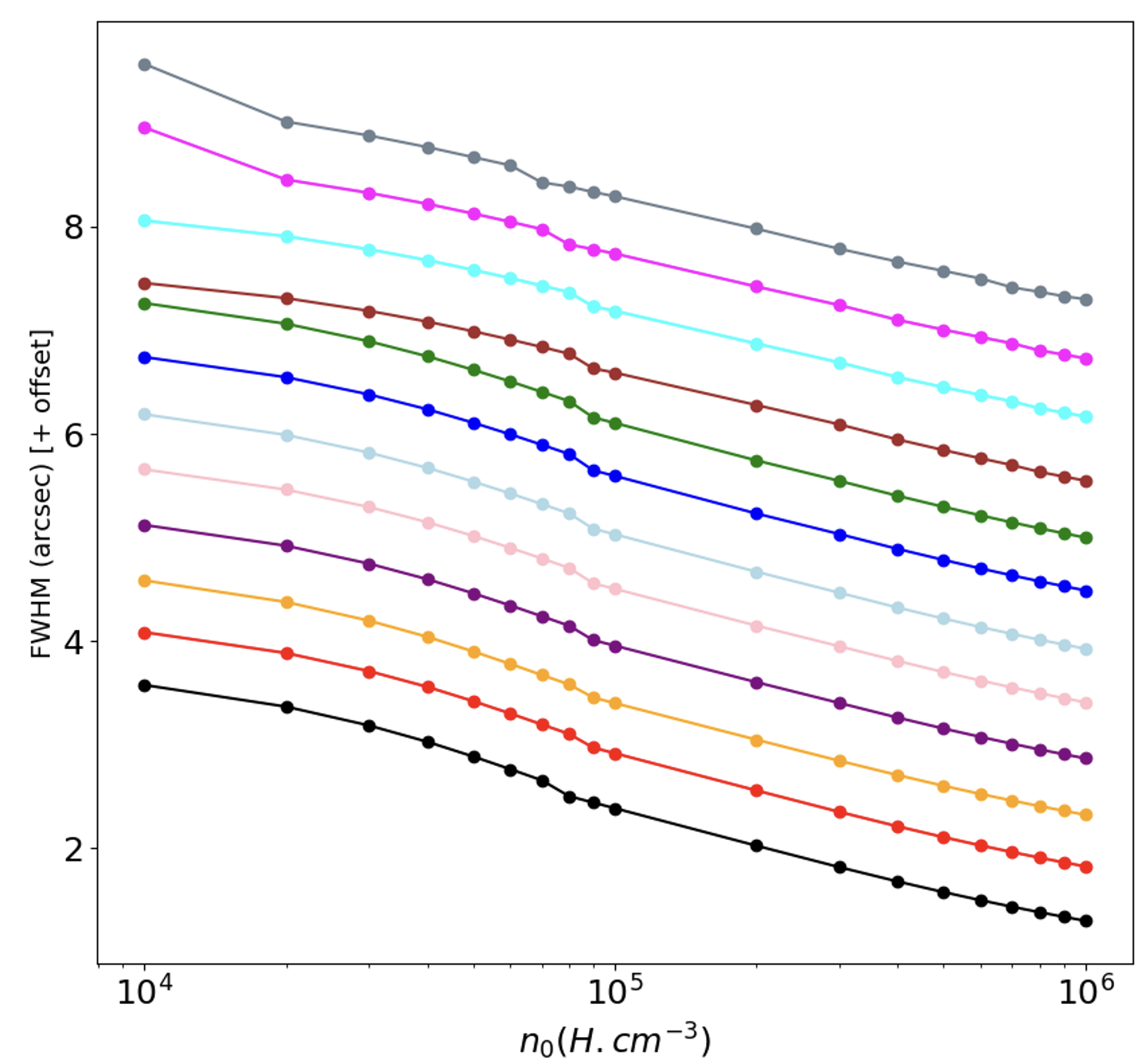}
    \includegraphics[width=0.455\linewidth]{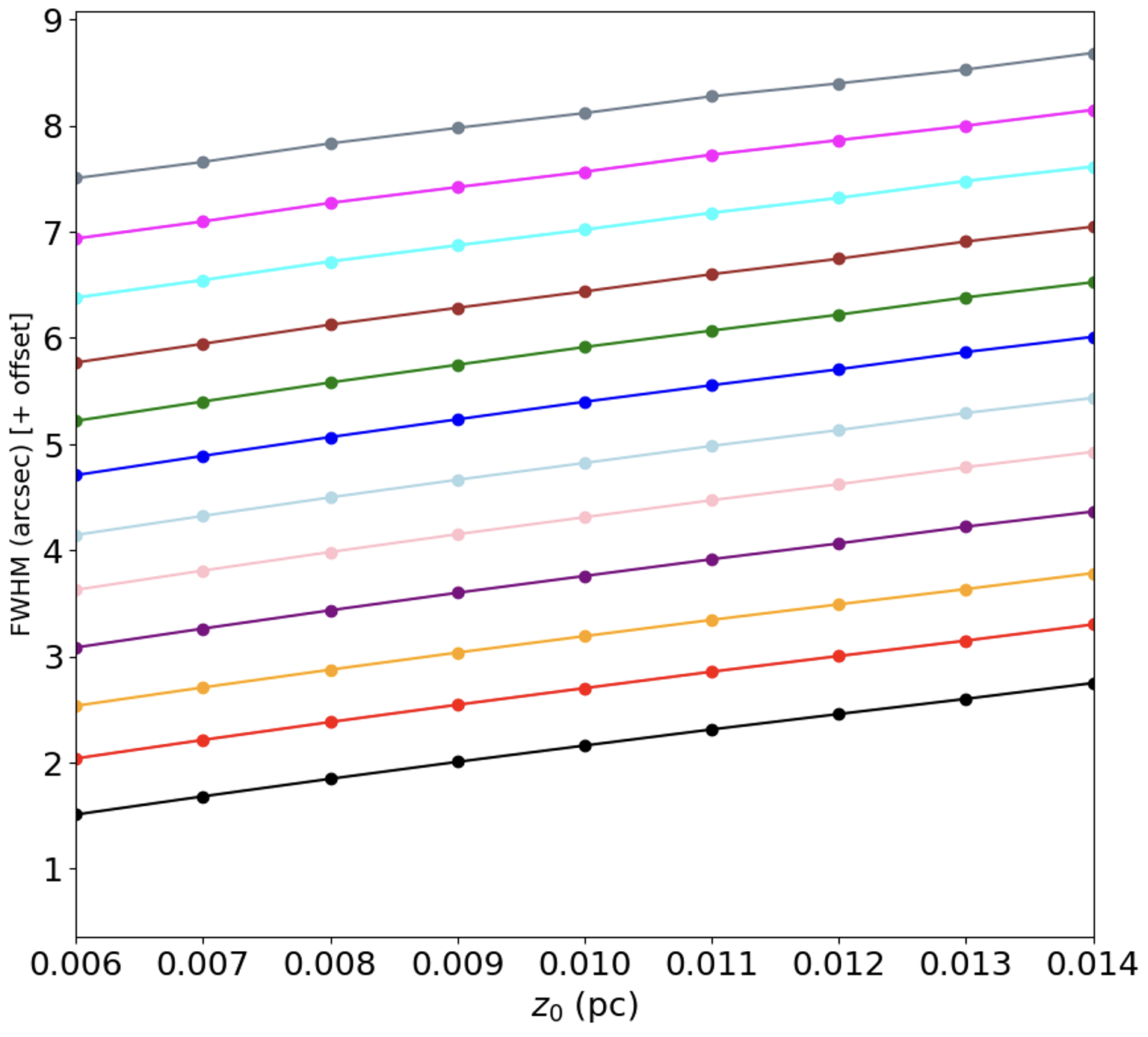}
    \caption{FWHM for each band as a function of nano-grains properties (\abvsg, \aminvsg, $\alpha$) and density profile (\(n_{0}\),\(z_{0}\)). Colours refer to the different photometric bands shown on the right (of the panels). An offset was applied for clarification (the lines are shifted from bottom to top in the order of increasing
wavelength).}\label{fig:fwhm_vs_dust}
\end{figure*}

\section{Comparison between JWST photometric and spectroscopic data and the self-coherent models using SOC and THEMIS. Different cases for $\alpha$ = -5}
\begin{figure*}
    \centering
    \includegraphics[width=\linewidth]{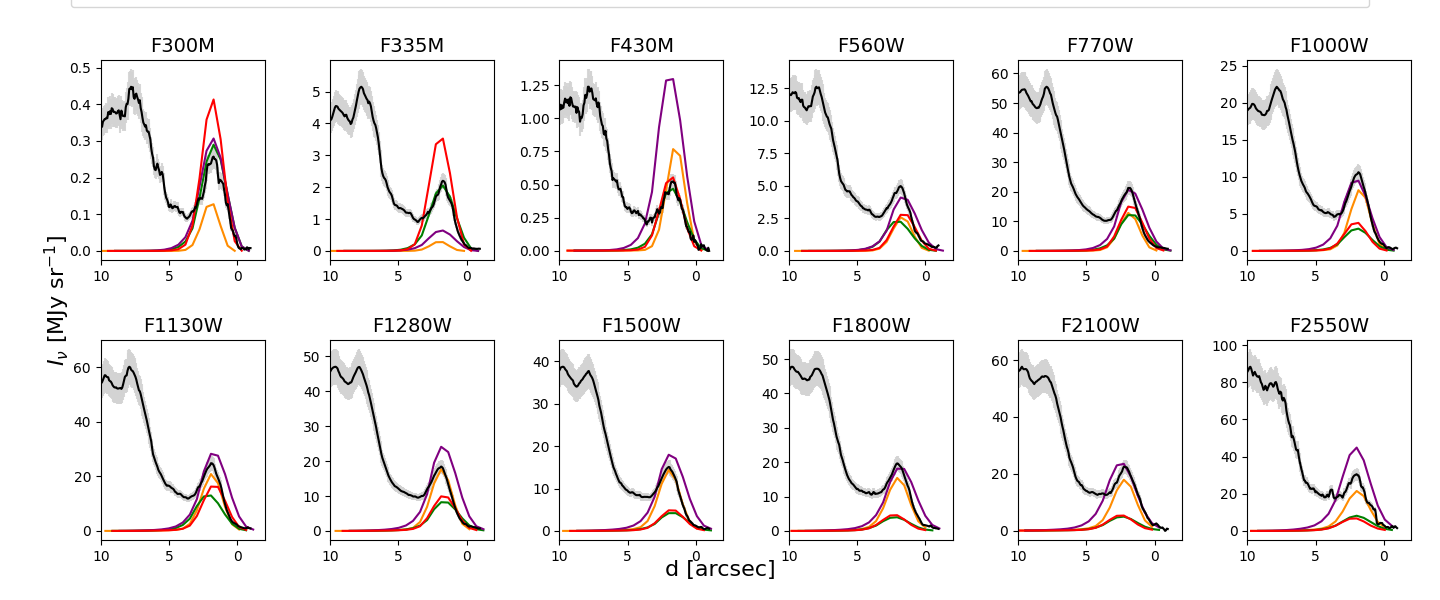}  \includegraphics[width=\linewidth]{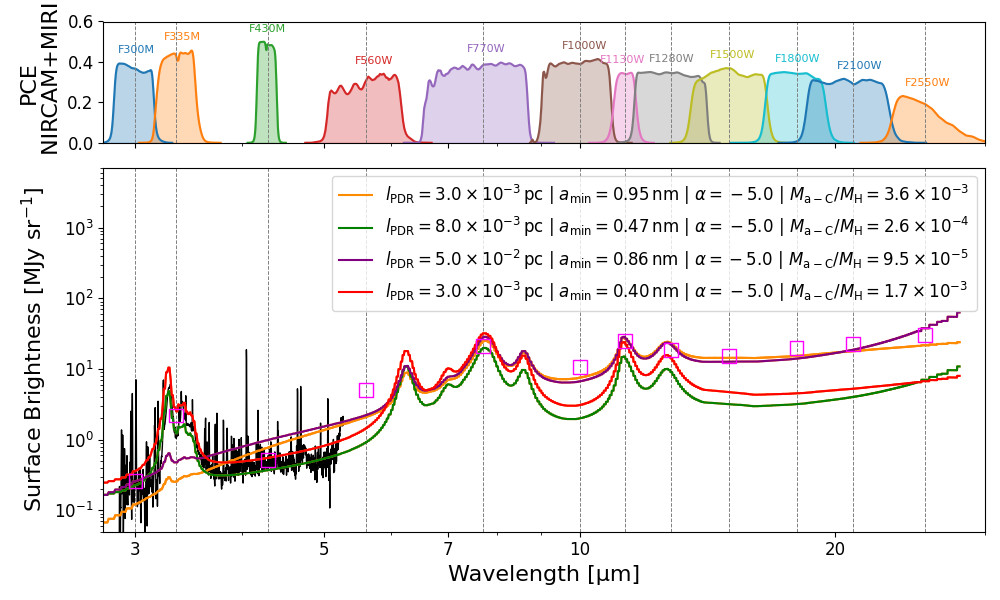}    
    \caption{Same as Fig. \ref{fig:best_profiles} but for $\alpha = -5$ with different values of $a_{\rm min}$: 0.95 nm (orange), 0.86 nm (purple), 0.47 nm (green), and 0.4 nm (red). For better visualization, we updated the photometric band colors and removed the MRS spectra. The maximum values of the photometric bands are marked with magenta squares.  }\label{fig:alpha-5}
\end{figure*}
\end{appendix}
\end{document}